\documentclass[manuscript,screen, nonacm]{acmart}

\usepackage{draftwatermark}
\SetWatermarkText{manuscript}
\SetWatermarkScale{0.8}
\SetWatermarkColor[gray]{0.9}

\setcopyright{none}
\renewcommand\footnotetextcopyrightpermission[1]{}
\AtBeginDocument{%
  }

\usepackage{hyperref}

\newcommand*{\eg}{e.g.,\@\xspace}
\newcommand*{\ie}{i.e.,\@\xspace}

\usepackage{subfigure}
\usepackage{verbatim}
\usepackage{multirow}
\usepackage{colortbl}
\usepackage{xspace} 
\usepackage{makecell}
\usepackage{stfloats} 
\usepackage{enumitem}
\usepackage{watermark}

\usepackage{xcolor}
\usepackage{color}
\usepackage{framed}
\usepackage{fancybox}

{\endMakeFramed}

\begin{document}

\title[What Clinicians Need]{What Clinicians Need: Designing, Developing and Evaluating an AI-Based Decision Support System for Autism Assessment}


\author{Ulrike Schäfer}
\authornote{Both authors contributed equally to this research.}
\orcid{0000-0002-9070-1665}
\affiliation{%
\institution{Freie Universität Berlin}
\department{Human-Centered Computing}
\city{Berlin}
\country{Germany}}
\email{ulrike.schaefer@fu-berlin.de}

\author{William Saakyan}
\authornotemark[1]
\orcid{0009-0002-9122-4722}
\affiliation{%
\institution{Bielefeld University}
\department{Human-Centered Artificial Intelligence}
\city{Bielefeld}
\country{Germany}}
\email{william.saakyan@uni-bielefeld.de}

\author{Matthias Norden}
\orcid{0000-0003-2377-524X}
\affiliation{%
\institution{Bielefeld University}
\department{Human-Centered Artificial Intelligence}
\city{Bielefeld}
\country{Germany}}
\email{mnorden@uni-bielefeld.de}

\author{Fabrizio Kuruc}
\orcid{0000-0002-3685-1005}
\affiliation{%
\institution{Freie Universität Berlin}
\department{Human-Centered Computing}
\city{Berlin}
\country{Germany}}
\email{fabrizio.kuruc@fu-berlin.de}

\author{Peter Sörries}
\orcid{0000-0003-0493-2895}
\affiliation{%
\institution{Freie Universität Berlin}
\department{Human-Centered Computing}
\city{Berlin}
\country{Germany}}
\email{peter.soerries@fu-berlin.de}

\author{Isabel Dziobek}
\orcid{0000-0003-0150-5353}
\affiliation{%
\institution{Humboldt-Universität zu Berlin}
\city{Berlin}
\country{Germany}}
\email{isabel.dziobek@hu-berlin.de}

\author{Claudia Müller-Birn}
\authornote{Both authors contributed equally to this research.}
\orcid{0000-0002-5143-1770}
\affiliation{%
\institution{Freie Universität Berlin}
\department{Human-Centered Computing}
\city{Berlin}
\country{Germany}}
\email{clmb@inf.fu-berlin.de}

\author{Hanna Drimalla}
\authornotemark[2]
\orcid{0000-0003-3783-7237}
\affiliation{%
\institution{Bielefeld University}
\department{Human-Centered Artificial Intelligence}
\city{Bielefeld}
\country{Germany}}
\email{drimalla@uni-bielefeld.de}

\renewcommand{\shortauthors}{Ulrike Schâfer, William Saakyan et al.}

\begin{abstract}

AI methods promise to support autism spectrum condition (ASC) diagnostics in adults, a complex and time-consuming process, that is characterized by a shortage of specialized clinicians. To date, clinicians' needs and their interaction with such AI-based support remain underexplored. Our work aims to develop and evaluate an AI-based clinical decision support system (CDSS) for ASC assessment, and to investigate how it impacts clinicians’ decision-making. By interviewing clinicians of varying experience levels, we identified five challenges and derived design strategies. Based on that, we developed SIT-CARE, a CDSS, which provides AI-based recommendations and data visualizations of clinically relevant nonverbal behavior. Through an evaluation study with newly recruited clinicians, we found that SIT-CARE led to different decision paths in regard to the ASC assessment, which are reflected in clinicians' mental models and decision changes. Overall, SIT-CARE demonstrated potential in improving initial diagnostic assessments, supporting in-depth diagnosis and empowering less experienced clinicians. 

\end{abstract}




\maketitle

\section{Introduction}
\label{sec:introduction}

Artificial Intelligence (AI) and Machine Learning (ML) have shown potential in improving diagnostic accuracy across various medical domains (\eg \cite{cai2019, laiScienceHumanAIDecision2023a}). 
In this work, we focus specifically on AI-assistance for the diagnosis process of Autism Spectrum Conditions (ASC) in adults.

ASC is recognized as a neurodevelopmental condition characterized by differences in social interaction and communication, as well as patterns of restricted or repetitive behavior~\cite{alma9918660570001341, Organization2004_ICD-10}.
According to the Centers for Disease Control and Prevention, approximately 3\% of the global population has ASC~\cite{shaw2025prevalence}.  
However, research indicates a high prevalence of undiagnosed ASC among adults with average or higher intelligence~\cite{lipinskiOutpatientPsychotherapyAdults2019, vogeleyDevelopmentSupportedEmployment2013}.
This may be due to the difficulty of diagnosing ASC in adults, particularly in those with robust coping or masking strategies\footnote{Individuals with ASC may use camouflaging or masking strategies to hide their autistic characteristics, with the intention of appearing ``non-autistic'' in social situations~\cite{mcquaidCamouflagingAutismSpectrum2022}.} ~\cite{fusar-poliMissedDiagnosesMisdiagnoses2022}.
ASC assessments currently in use are time-consuming and require extensive training, but they only have moderate accuracy on their own~\cite{Conner2019_, Ashwood2016_utilityAQ10} and often rely on patients' subjective self-reports~\cite{Zhu2025_AIDoctorforASD}. 
Furthermore, due to a shortage of specialized clinicians~\cite{lipinskiBlindSpotMental2022, volkmar2022diagnostic}, the high demand for ASC diagnostics cannot be met~\cite{davidsonNoExclusionsDeveloping2015}.
%
To address these challenges, research has begun to explore the potential of using AI to support ASC assessments (see~\cite{joudarArtificialIntelligencebasedApproaches2023, kleineAIenabledClinicalDecision2025}). 
One promising approach is training an AI on more objective, behavioral data~\cite{simeoliUsingMachineLearning2024}. However, such data needs to be collected under standardized conditions.
For this purpose, the Simulation Interaction Task (SIT)~\cite{drimallaAutomaticDetectionSocial2020} can be used.
The SIT is a standardized task that supports the assessment of adults with ASC. During the task, patients interact with a pre-recorded actress via video about their favorite and least favorite foods. This enables the collection of video data during the interaction. 
These SIT videos can be used to extract nonverbal behavioral data from patients and train AI models, as prior work has shown~\cite{drimallaAutomaticDetectionSocial2020}.

However, recent research suggests that assisting decision-making with AI rarely leads to improved human-AI team performance, underlining the need to develop human-centered AI (HCAI) that involves users in the design process~\cite{capelWhatHumanCenteredHumanCentered2023, vaccaroWhenCombinationsHumans2024}. 
%
Thus, to support ASC assessment with AI, clinicians' needs and their workflows should be considered, which leads to our first research question: ``\textbf{What challenges do clinicians face in the decision-making process of diagnosing ASC in adults, and how can AI support them?} (RQ1)''.

To accommodate clinicians and their workflow appropriately with a human-centered AI-based CDSS, in addition to considering their needs and the type of task (\eg decision-making~\cite{vaccaroWhenCombinationsHumans2024, hemmer2021}), clinicians need to be involved in evaluating the system~\cite{capelWhatHumanCenteredHumanCentered2023}. Especially in the medical context, clinicians' decision-making process needs to be investigated thoroughly~\cite{wangHumancenteredDesignEvaluation2023, sox2024medical}.
This leads to our second research question, in which we investigate how an AI-based clinical decision support system (CDSS) for ASC assessment in adults influences clinicians' decision-making processes including their mental models of the AI\footnote{A mental model is a simplified and incomplete understanding of how an AI system works and behaves~\cite{norman1988psychology, kulesza2012, bansal2019beyond}.}:
%
``\textbf{How does an AI-based CDSS for ASC assessment influence clinicians' decision-making processes and mental models?} (RQ2)''


To answer our first research question (RQ1), we interviewed clinicians of varying levels of experience to gain insight into their needs and diagnostic workflows. 
These interviews revealed that clinicians want AI support, but five challenges were identified, each with its own design strategy for implementation. These include, for example, what and how additional information should be provided by considering current workflows.
To the best of our knowledge, there is currently no AI-based CDSS for assessing ASCs using standardized interactions, such as the SIT.
\textit{To bridge this gap, we developed and evaluated SIT-CARE that supports ASC assessment in adults, following the identified design strategies.}
SIT-CARE uses more objective, nonverbal behavioral data extracted under standardized conditions from video and audio recordings of the SIT. 
As previous HCI research has shown that clinicians may need support in different stages of the medical decision-making process~\cite{zhangRethinkingHumanAICollaboration2024, wuCardioAIMultimodalAIbased2025}, \ie not only in the final stage (\textit{making decisions}), but also in earlier stages (\textit{generating hypotheses}, \textit{gathering data}, \textit{testing hypotheses})~\cite{sox2024medical, zhangRethinkingHumanAICollaboration2024, wuCardioAIMultimodalAIbased2025}, we considered this aspect in our development.
Based on the SIT data, we generated visualizations and numerical summaries of three modalities, \ie gaze behavior, facial expressions, and voice parameters. 
These visualizations and numerical summaries are provided to clinicians for a \textsc{data-based assessment} to support them in earlier stages of decision-making, such as generating hypotheses and gathering data~\cite{sox2024medical}.
%
Additionally, we trained AI models using nonverbal data beyond the three modalities to provide \textsc{model-based assessment}. This \textsc{model-based assessment} provides an AI recommendation on whether to perform further ASC diagnostics, accompanied by information about the model and prediction to support later decision-making stages, \ie making the decision~\cite{sox2024medical, zhangRethinkingHumanAICollaboration2024}.

To answer our second research question (RQ2), we conducted an evaluation study to investigate how SIT-CARE influences clinicians' decision-making. 
In this study, clinicians assessed two patient cases, one with ASC and one without (Non-ASC), three times: First based on the SIT video, then again after gathering information and exploring the visualizations in the \textsc{data-based assessment}, and finally, after receiving the AI recommendation in the \textsc{model-based assessment}.
We identified different decision paths that are reflected in the clinicians’ mental models and characterized by changes in decisions after receiving information from SIT-CARE. 
Interestingly, even when the clinicians followed the same decision path, their reasoning and mental models differed.
For example, some participants only considered the \textsc{model-based assessment} and did not see the need to understand the AI's reasoning, while others studied the \textsc{data-based assessment} in detail.
Additionally, all clinicians found SIT-CARE to be valuable for ASC screening support\footnote{A screening is a brief initial assessment before an in-depth diagnosis is considered.}, and also as a valuable learning opportunity.

This paper provides the following contributions:
\begin{itemize}
    \item Through a formative study with clinicians, we identified five challenges and derived design strategies for developing an AI-based CDSS for ASC assessment in adults.
    \item We developed a new AI-based CDSS for ASC assessment, SIT-CARE, which supports clinicians in analyzing patient's nonverbal behavior, and gives an AI recommendation whether further ASC assessment may be needed. 
    \item In an evaluation study, we identified different decision paths among clinicians, which are reflected in their mental models and decision changes. Clinicians expressed high interest in our system for screening support and as a learning opportunity.
\end{itemize}

\section{Related Work}

\subsection{Autism spectrum condition diagnostics in adults}
\label{subsec:RW_ASC_AI_Assistance}

Currently, ASC diagnostics rely on behavioral and observational assessments due to the lack of conclusive biomarkers~\cite{Zhu2025_AIDoctorforASD}.
For a brief ASC screening, questionnaires such as the Autism Questionnaire (AQ \cite{baron-cohenAutismSpectrumQuotientAQ2001}) and the Ritvo Autism and Asperger Diagnostic Scale-Revised (RAADS-R \cite{Ritvo2011_TheRitvoAutismAspergerDiagnosticScale-RevisedRAADS-R}) are used to flag adults for referral to a specialist. However, these rely on self-reporting and their utility is disputed \cite{Ashwood2016_utilityAQ10, Conner2019_, Jones2021_}.
Thus, the current standard for diagnosing adults with ASC is a semi-structured assessment, \ie the Autism Diagnostic Observation Schedule (ADOS~\cite{lordAutismDiagnosticObservation}), a clinician-administered observational protocol in which social, communicative, and behavioral responses are elicited through structured tasks, or the Autism Diagnostic Interview–Revised (ADI‑R~\cite{Lord1994_AutismDiagnosticInterview--Revised}), which is conducted by a specialist and does not rely solely on self-reporting. 

Nevertheless, these instruments are time-consuming, require extensive training, and only have moderate accuracy in adults on their own~\cite{Conner2019_}. 
\citet{Conner2019_} found that even the gold standard ADOS had only \~65\% sensitivity in an outpatient sample, indicating the difficulty of diagnosing ASC in adults.
In addition, adults with higher intelligence often develop coping and masking strategies that can complicate clinicians' assessment~\cite{fusar-poliMissedDiagnosesMisdiagnoses2022, lipinskiOutpatientPsychotherapyAdults2019, vogeleyDevelopmentSupportedEmployment2013}.
%
%
These challenges of traditional methods led to a growing interest in using AI to support and improve the ASC diagnostic and screening process.
%

\subsection{AI-supported diagnostics for autism spectrum conditions}
\label{subsec:RW_ASC_AI_SIT}

AI has been applied to adult ASC screening questionnaires, such as the AQ-10 and RAADS-R (\eg \cite{Thabtah2019_ASCtest}). 
However, training AI with subjective self-reported experiences can be problematic, as these reports can be unreliable or incomplete, particularly in adults who use coping or masking strategies.
Other AI-based approaches focus on behavioral information similar to current diagnostic assessments like the ADOS.
For example, \citet{Kumar2025_PredictingAutismSpectrumDisorderinAdultsThroughFacialImageAnalysis} have used AI techniques to identify subsets of behavioral features from the ADOS that achieved high sensitivity and specificity, comparable to the AI model considering all ADOS features. 
Another promising approach is the automated analysis of standard video recordings~\cite{Syriopoulou2025}. 
Such AI models have shown strong potential in detecting subtle and complex behavioral cues, such as facial expressions \cite{Briot2021, Kumar2025_PredictingAutismSpectrumDisorderinAdultsThroughFacialImageAnalysis}, gaze dynamics \cite{Kim2024_}, and gesture patterns~\cite{lakkapragadaClassificationAbnormalHand2022} that may signal autistic traits and could be missed by standard clinical observation.

Currently, few AI approaches for ASC in adults have progressed from experimental validation to practical, scalable diagnostic tools \cite{Cavus2021_}. 
%
Thus, there is a need for AI-based CDSS that can be integrated into routine procedures for ASC diagnostics, enabling structured, multimodal observation of social-communicative behavior in adults.
To train such an AI model, social-communicative behavior must be collected under standardized conditions.
For this, the \textit{Simulated Interaction Task (SIT) }\cite{drimallaAutomaticDetectionSocial2020}, a computer-based social interaction paradigm designed to elicit verbal and nonverbal behavior under standardized conditions in adults, can be used. 
For example, previous research already showcased the scalability and accessibility of the SIT by collecting large datasets (\eg ~\cite{drimallaAutomaticDetectionSocial2020, Saakyan2023_a}). 
%
Further, these studies demonstrated that using these datasets, consisting of video and audio data, a multimodal AI model can be trained by extracting facial expressions, head movements, and vocal features, which distinguished adults with and without an ASC diagnosis (accuracy > 70\%).

To date, such an AI model trained on more objective, standardized interaction data of the SIT has not yet been integrated into a human-centered application for clinicians to support ASC assessments.

\subsection{Designing and developing clinician-centered AI}
\label{subsec:RW_constructs}

Our goal of providing clinicians with an AI-based CDSS is to support their decision-making with respect to ASC assessments in adults.
However, a recent meta-analysis found that human-AI teams, \ie humans assisted by AI, do not outperform their own or the AI's performance in decision-making tasks~\cite{vaccaroWhenCombinationsHumans2024}.
Highly researched approaches that ``\textit{produce details or reasons to make [AI's] functioning clear or easy to understand}'', \ie Explainable AI (XAI)~\cite{barredoarrieta2020}, did not lead to improved team performance. 
This could be due to the technology-focused perspective ``\textit{aiming to enhance algorithmic effects rather than meet user needs}''~\cite{Zhu2025_AIDoctorforASD}.
Hence, the field of Human-Computer Interaction (HCI) has begun to shift to human-centered design approaches in order to develop HCAI~\cite{capelWhatHumanCenteredHumanCentered2023}. 
HCAI encompasses involving humans throughout the process, from design to evaluation. In addition, aspects such as the type of task should be considered more~\cite{vaccaroWhenCombinationsHumans2024, hemmer2021}.

Especially in the medical field, as highlighted by
\citet{capelWhatHumanCenteredHumanCentered2023}, human-centered approaches are often used by involving clinical specialists in the development process and considering the context of the interaction, such as the medical decision-making workflow (\eg \cite{zhangRethinkingHumanAICollaboration2024, wolfHowClinicalDecision2025, wuCardioAIMultimodalAIbased2025, rosbachWhenTwoWrongs2025, wang2019, wangBrilliantAIDoctor2021}). 
For example, \citet{zhangRethinkingHumanAICollaboration2024} conducted a formative study involving clinicians to improve an existing AI-powered sepsis prediction module, followed by an evaluation study. 
By considering the clinicians' decision-making process, the authors found that they need to support earlier medical decision-making stages (\eg hypotheses generation and data gathering~\cite{sox2024medical}) instead of only focusing on later stages (\eg final decision~\cite{sox2024medical}).
Similarly, \citet{wolfHowClinicalDecision2025} focused on earlier medical decision-making stages, and found that their AI-based CDSS triggered analytical thinking.

Thus, to design and develop a \textbf{human-centered AI-based CDSS}, 
clinicians need to be involved throughout the design process~\cite{capelWhatHumanCenteredHumanCentered2023} and their needs, the type of task~\cite{hemmer2021, vaccaroWhenCombinationsHumans2024}, in this case medical decision-making steps~\cite{sox2024medical}, have to be considered.

\subsection{Understanding users' mental models during AI-assisted decision-making}
\label{subsec:RW_Mental_Models}

To understand how clinicians use such an implemented AI system and improve their interaction with it, we need to investigate the clinicians' reasoning during their interaction with the AI system, \ie their \textit{mental model}.
A mental model is a simplified, incomplete understanding of how the AI behaves and works that a user develops as they interact with an AI system~\cite{norman1988psychology, kulesza2012, bansal2019beyond}. This understanding influences how users interact with, comprehend, and predict the AI system's actions~\cite{norman2014some, kelly2023a, hoffmanMeasuresExplainableAI2023}. 
Studies indicate that a user's mental model of an AI system influences their ability to interact with it effectively, which in turn may affect the performance of human-AI teams~\cite{cabrera2023, bansal2019beyond}, for example, diagnostic accuracy.
Various methods can be used to explore a user's reasoning and mental model of an AI system, ranging from questionnaires (\eg \cite{lai2022, cheng2019a, kulesza2012}), task performance and behavior (\eg \cite{bansal2019beyond, narayanan2023}) to prediction tasks~\cite{hoffmanMeasuresExplainableAI2023}.
Qualitative methods like think-aloud protocols and interviews can investigate a user's mental model in more depth, revealing why they made a decision (\eg \cite{gero2020, cabrera2023}).
Furthermore, research investigated users' reliance on AI by considering different decision paths that humans could take~\cite{schemmer2023appropriate, morrisonImpactImperfectXAI2024}. 
For example, a user may be asked to make decisions at different \textit{decision points}, such as before and after receiving AI advice and explanations. 
Exploring which decision paths were taken and when changes occurred could help us better understand a user's decision-making process.

To support ASC assessment, it is important that the AI-based CDSS used accommodates clinicians appropriately. Therefore, their decision-making process during interacting with the CDSS has to be investigated in detail~\cite{wangHumancenteredDesignEvaluation2023, yangUnremarkableAIFitting2019}. 
To accomplish this, we explored the mental models of users by combining two of the aforementioned approaches.
First, to investigate the clinicians' decision-making process in more detail, we asked the clinicians to decide on several decision points instead of letting them make just one decision after receiving the AI recommendation~\cite{bansal2019beyond, hoffmanMeasuresExplainableAI2023}. This may allow us to identify the decision paths clinicians took. For example, it could show that AI-based information only leads to decision changes under specific circumstances.
To investigate why specific decision paths were taken and why decisions may have changed, qualitative methods (\eg \cite{bachIfHadAll2023, zhangRethinkingHumanAICollaboration2024, wuCardioAIMultimodalAIbased2025}) such as the thinking-aloud method can be used. These methods allow us to ask questions at each decision point to gain an in-depth understanding of a user’s current mental model and reasoning.

By studying clinicians' mental models and reasoning during their decision-making process in more detail, we can gain a richer understanding of how our AI-based CDSS influences their decisions.

\begin{table*}[!htbp]
    \captionsetup{justification=centering}
    \small{
    \caption{Demographics of Participants in our Formative Study.}
    \Description{This table presents the demographics of participants in our formative study. The table has 4 columns: P#: A unique identifier for each participant. Gender: The gender of each participant. Professional Stage: The stage of professional development of each participant, which can be either a psychotherapist in training or a practicing psychotherapist. Experience: The amount of experience each participant has in their profession, including the number of years and whether they are specialized in ASC. The table has 7 rows, each representing a different participant, P1 to P7. Here are the values listed per row P#: P1 is a female psychotherapist in training with 0.5 years of experience. P2 is a female practicing psychotherapist with 5 years of experience, specialized in ASC. P3 is a female practicing psychotherapist with more than 15 years of experience, specialized in ASC. P4 is a female psychotherapist in training with 3 years of experience. P5 is a female practicing psychotherapist with 25 years of experience, specialized in ASC. P6 is a female practicing psychotherapist with 5 years of experience, specialized in ASC. P7 is a female practicing psychotherapist with more than 25 years of experience, specialized in ASC.
    }
    \label{tab:demographics_1}
    \renewcommand{\arraystretch}{1.5}
    \begin{tabular}{ccp{6cm}p{5cm}}
        \Xhline{3\arrayrulewidth}
        \textbf{P\#} & 
        \textbf{Gender} & 
        \textbf{Professional Stage} & 
        \textbf{Experience}
        \\ 
        \Xhline{2\arrayrulewidth}
        P1 & 
        Female & 
        Psychotherapist in Training & 
        0.5 years
        \\ 
        P2 & 
        Female & 
        Practicing psychotherapist & 
        5 years, specialized in ASC \\
        P3 & 
        Female & 
        Practicing psychotherapist & 
        >15 years, specialized in ASC \\
        P4 & 
        Female & 
        Psychotherapist in Training & 
        3 years\\
        P5 & 
        Female & 
        Practicing psychotherapist & 
        25 years, specialized in ASC\\
        P6 & 
        Female & 
        Practicing psychotherapist
        & 
        5 years, specialized in ASC\\
        P7 & 
        Female & 
        Practicing psychotherapist & 
        >25 years, specialized in ASC\\ 
        \Xhline{3\arrayrulewidth}
    \end{tabular}%
    }
\end{table*}

\section{Formative Study: Challenges and design strategies for AI-assisted ASC assessment support}
\label{sec:study1}

To guide our design and development of an AI-based CDSS for ASC assessments, we conducted open-ended semi-structured interviews~\cite{longhurst2003semi}. 
Through these interviews, we aimed at exploring clinicians' needs and challenges they face during the decision-making process of diagnosing ASC in adults.
Further, we gathered feedback on how to effectively communicate nonverbal behavior extracted from video recordings of individuals during the SIT, \ie a standardized social interaction paradigm (see \autoref{subsec:RW_ASC_AI_SIT}).

\subsection{Method}

We recruited seven clinicians, \ie practicing psychotherapists, via purposive sampling~\cite{andradeInconvenientTruthConvenience2021}, see \autoref{tab:demographics_1}. 
To understand the clinicians' needs, we included participants with varying levels of expertise, ranging from psychotherapists in training to practicing psychotherapists who specialize in ASC.
Three of the authors were present for each interview, which was conducted remotely via an institutional Webex platform. Each session lasted an average of 49 minutes.
This study was approved by the IRB of the second author's institution (application number, institution, and date will be provided after acceptance). All participants provided informed consent before participating.

During the interview, participants were asked questions about their demographic background, their experience with ASC, and their diagnostic workflows. 
Next, we introduced the SIT and explained that video data of a person during the SIT can be recorded and used to analyze nonverbal behavior.
We then asked participants to identify nonverbal behavior they considered most relevant for ASC diagnosis.
After capturing their initial thoughts, the clinicians were shown a list of nonverbal behaviors and asked to assess which of these behaviors would be helpful based on the provided information.
We presented four schematic visualizations of nonverbal behavior, including gaze, facial expressions, voice, and head behavior (see \autoref{sec:app_interviewscript_1}), to gather feedback on their usefulness and integration.
By showing examples with different plot types, \ie line, box, and bi-dimensional histogram plots, we were able to get feedback on what plot type is the most useful, liked, and on how such information could be integrated into an implementation.
Finally, we asked about what challenges and benefits of integrating SIT and its data visualizations into their diagnostic workflows they expect.
Details of the interview protocol are provided in \autoref{sec:app_interviewscript_1}.

A total of six hours of audio material was recorded, transcribed, and analyzed using an inductive coding approach~\cite{mayring2014content}. We read the transcripts multiple times to ensure familiarity, and derived the initial codes by segmenting the text into meaningful units, which were discussed in detail by all three interviewers. These codes were iteratively grouped into higher-level categories and refined into overarching themes. 
The findings were then collaboratively structured and refined through iterative feedback and team discussions.

\subsection{Findings}
\label{subsec:study1_findings}

Our interviews revealed five challenges and led to five design strategies for an AI-based CDSS for ASC screening.

\subsubsection{Supporting clinicians to meet diagnostic demand}
\label{subsec:study1_empower}

All clinicians underlined that diagnosing ASC requires years of training and experience compared to other diagnoses, as knowledge about specific alterations of nonverbal behavior in affected individuals is necessary (P5, P2, P7). 
Furthermore, P5 described ``\textit{We open up for new registrations every three months, [...] we had 1,400 registrations for an estimated 30 people that we can actually take on.}''
%
Consequently, due to the high demand for autism diagnoses, there has been a need to support clinicians in their assessments and enable more clinicians to perform initial screenings (P6, P2).

To address this, P5 recommended ``\textit{to outsource certain parts of the diagnostic process}'' and noted that ``\textit{nonverbal symptoms are something that clinicians who are not familiar with the field find particularly difficult.}''
P2 commented that ``\textit{[AI-assisted support] could be really cool in a screening context, because then people [...] can be very well-supported in this case, and maybe even identify people who can continue with the diagnostic process.}''.
All participants expressed interest in AI-assisted support, but noted the need for additional guidance, such as definitions for clinical and technical terms. (P1).

In summary, due to the high demand for autism diagnosis, clinicians need additional support, such as AI-based assistance, to improve both the screening and diagnosis process and empower less experienced clinicians \textit{(Design Strategy 1)}.

\subsubsection{Providing understandable information to foster more objective ASC assessments}
\label{subsec:study1_comm}

Participants raised the concern that the assessments currently used to diagnose ASC heavily rely on patients' self-reports and clinicians' interpretations, as no more ``\textit{objective}'' measures, such as biomarkers, are currently established (P1, P3, P6). 
%
P7 explained, ``\textit{[The ADOS] is not very informative [especially for high functional ASC], and in the end it is just a subjective assessment by us [the clinicians]}''. 
%
P2 reflected: ``\textit{Do I really recognize well, even though I've been doing this for a bit longer now, whether this person is actually very, very good at masking or if they're just good at [social interactions]?}'' and explained that gender bias may influence the clinician's subjective assessment.
It is also challenging that most patients already suspect they have ASC and may disagree with a clinician's assessment (P3).
In such cases, clinicians often feel pressured to provide more objective evidence (P7).

Participants emphasized that ``\textit{[to make the decision] less dependent on my subjective impression, but rather to make it somehow more objective, [...] would be fantastic}'' (P2) and a ``\textit{a great additional proof.}'' (P7).
P4 mentioned that they would carefully select some visualizations representing nonverbal behavior to discuss them with patients. 
Summarized, all participants expressed interest in visualizations of nonverbal behavior, as long as they are understandable.
By collecting feedback on the example visualizations, we were able to extract how the participants want more objective data to be presented.
While box plots were described as not helpful or understandable (P2, P3, P6), line plots showing information over time were preferred (P1, P2, P4). The bi-dimensional histogram plot for gaze information was described as intuitive (P2, P6, P7), but it was recommended to illustrate the display and actress in the plot (P5).

Summarized, participants want more objective information presented in an understandable way \textit{(Design Strategy 2)}.

\subsubsection{Balancing the amount and detail of information}
\label{subsec:study1_less}

During the interviews, we asked clinicians which nonverbal behavioral data would be of interest and presented visualization drafts. 
While all clinicians expressed high interest in most of the data, it was also noted that providing too much information could lead less experienced clinicians to consider irrelevant aspects, as P5 states: \textit{``it's always, a bit less is more''}. Also, the worry that providing such an analysis of nonverbal behavior may be too complicated was mentioned (P6).

The participants recommended to focus on nonverbal behavior that \textit{``have been found to be the most informative in research and clinicians also observe themselves''} or \textit{``that are in the }ICD''\footnote{ICD is the \href{https://icd.who.int/en/}{International Classification of Diseases}.} (P5).  
Additional details could be presented in tables or numerical formats (P5, P7).
P6 emphasized the importance of detail on demand, stating \textit{``so that I can get important information at a glance, but more information if I want it.''} Similarly, P4 suggested \textit{``I would show all the visualizations, but not all at once [...]. You can unfold them individually, so that you're not overwhelmed by so many presentations.''}, while also recommending the ability to \textit{``interactively switching between the reference groups''.}

In summary, nonverbal behavior should be selected based on clinical relevance, with adjustable information levels (\textbf{Design Strategy 3}).

\subsubsection{Considering the existing clinical workflow}
\label{subsec:study1_familiar}

In the interviews, it became apparent that the diagnostic process is ``\textit{extremely time-consuming}'' (P2, P6).
Similarly, P1 explicitly noted that the current ASC diagnostics already consists of many different assessments and that integrating anything new into their workflow would be difficult. 
%
For example, the participants described that besides many other assessments, they also judge a person's nonverbal behavior with the ADOS (P3, P6, P7). Sometimes, even case conferences, supervisions, or feedback from colleagues may be needed, especially for edge cases (P2, P4, P6). 

Participants stated that many of the discussed nonverbal behaviors are similar to aspects they are familiar with from the ADOS, and that data-based information would be a good addition (P2, P6).
P4 and P6 highlighted the advantage of having standardized video recordings, which could be discussed in case conferences.
Furthermore, multiple participants underlined that their expertise is built on a lot of training and experience from numerous cases (P2), and that providing visualizations of example cases as an orientation would be helpful (P5, P7).
In addition, participants want to be able to compare current patient data with reference values, such as from control groups without ASC  (P2, P5), differentiated by gender or other diagnoses (P2, P7). 
Similarly, P5 described that \textit{``[they] would like to see the reference values for people without autism in the image [...]  with standard deviations or something like that''}, which is similar to their current practice of comparing a patient's questionnaire score with references.

Summarized, for a seamless integration into existing diagnostic workflows, we aim to build on concepts that are familiar to clinicians \textbf{ (Design Strategy 4)}.

\begin{table}[htbp]
    \captionsetup{justification=centering}
    \small{
    \caption{Challenges, Design Strategies and Implementation based on the Formative Study}
    \Description{This table presents the challenges, design strategies, and implementation findings from the formative study, which are described in detail in section 3.2. The table has 3 columns: Challenge: The challenges faced by clinicians in the diagnostic process for ASC. Design Strategy: The design strategies proposed to address the challenges faced by clinicians. Summary of clinicians' needs and wants: A summary of the needs and requirements identified by clinicians to support the diagnostic process.
    The table has 5 rows, each representing a different challenge, design strategy and a description of clinician's specific needs and requirements. Here are the values listed per row: First row: The high demand for autism diagnosis is a challenge, and the design strategy to address this is to provide additional support in the diagnosis process. Clinicians' needs include: Support and education on judging nonverbal behavior, Outsourcing parts of the diagnostic process and AI-assisted support in ASC screening wanted. The second row: Current assessments are subjective impressions of clinicians and influenced by the patient's own impression. The design strategy to address this is to provide understandable information to foster objectivity in ASC assessments. Clinicians' needs include: Present understandable analyses and visualizations of nonverbal behavior, Intuitive plots (e.g. line plot, bi-dimensional histogram plot) are preferred over box plots, Gaze behavior towards the display can be illustrated directly on an illustration with a display and the actress. The third row: The current assessments are too much information may be overwhelming. The design strategy to address this is to balance the amount and detail of information. Clinicians' needs include: Only selected nonverbal behavior should be visualized, Additional information can be provided in text or tables, Allowing user to control the wanted amount of information with an interactive interface. The fourth row: The already time-consuming diagnosis process is a challenge, and the design strategy to address this is to consider the existing clinical workflow. Clinicians' needs include: Connect the new data-based information with known concepts (e.g. ADOS), Provide visualizations depicting nonverbal behavior from example cases as a comparison Provide and visualize reference values. The fifth row: The misuse of provided recommendations and information is a challenge, and the design strategy to address this is to set expectations. Clinicians' needs include: Communicate intended use and limitations clearly, Inform about the system's accuracy.
    }
    \label{tab:tab_study1_findings}
    \renewcommand{\arraystretch}{1.5}
    \begin{tabular}{p{3.5cm}p{3.5cm}p{7cm}}
        \Xhline{3\arrayrulewidth}
        \textbf{Challenge} & 
        \textbf{Design Strategy} & 
        \textbf{Summary of clinicians' needs and wants}
        \\ \Xhline{1\arrayrulewidth}
        High demand for autism diagnosis & 
        Additional support in their diagnosis process & \vspace{-\baselineskip}
        \begin{itemize}[leftmargin=*,label=\textendash]
          \item Support and educate about judging nonverbal behavior
          \item Outsourcing parts of diagnostic process
          \item AI-assisted support in ASC screening wanted
        \end{itemize} 
        \\ 
        Current assessments are subjective impressions of clinicians and influenced by the patient's own impression & 
        Providing understandable information to foster more objective ASC assessments & 
        \vspace{-\baselineskip}
        \begin{itemize}[leftmargin=*,label=\textendash]
          \item  Understandable 
        visualizations of nonverbal behavior  
          \item  Intuitive plots (\eg line plot,  bi-dimensional histogram plot) are preferred over box plots 
          \item Gaze behavior towards the display can be illustrated directly on an illustration with a display and the actress 
        \end{itemize} 
        \\ 
        Too much information may be overwhelming  & 
        Balancing the amount and detail of information & 
        \vspace{-\baselineskip}
        \begin{itemize}[leftmargin=*,label=\textendash]
            \item   Only selected nonverbal behavior should be visualized  
            \item Additional information can be provided in text or tables 
            \item Allowing user to control the wanted amount of information with an interactive interface
        \end{itemize}
        \\ 
        Already time-consuming diagnosis process & 
        Considering the existing clinical workflow & 
        \vspace{-\baselineskip}
        \begin{itemize}[leftmargin=*,label=\textendash]
          \item   Connect the new data-based information with known concepts (\eg ADOS) 
          \item Provide visualizations depicting nonverbal behavior from example cases as a comparison 
          \item Provide and visualize reference values 
        \end{itemize}
        \\ 
        Misuse of provided recommendations and information & 
        Setting Expectations & 
        \vspace{-\baselineskip}
        \begin{itemize}[leftmargin=*,label=\textendash]
          \item  Communicate intended use and limitations clearly 
        \item Inform about the system's accuracy 
        \end{itemize}
        \\
        \Xhline{3\arrayrulewidth}
    \end{tabular}%
    }
\end{table}

\subsubsection{Setting expectations}
\label{subsec:study1_limitations}

Participants critically reflected on how such a tool should be introduced and how uncertainty should be communicated.
Participants explained that \textit{``this little building block [data-based support] is intended to help with nonverbal communication, this small part of the diagnosis, and to do so with a certain level of accuracy''} (P5).
It should be avoided to diagnose  \textit{``based on a single symptom''} (P7).
P5 further stated that \textit{``The responsibility also lies with the developers'}' to ensure that a tool is used as intended.

%
P7 recommended to ``\textit{exactly explain [to the clinicians] what this little building block is [made] for}'', and that it does not replace the diagnosis process (P5).
Furthermore, participants expressed the need to know the system's accuracy, for example, the probability of the system recognizing the nonverbal behavior as typical for ASC (P5). 

Summarized, users need to be informed about the intended use and system limitations (\textbf{Design Strategy 5}).

\subsection{Summary of Findings}

Answering our first research question, our formative study revealed challenges clinicians face in their current ASC diagnoses, as well as design strategies that could address these challenges, see \autoref{tab:tab_study1_findings}.
Currently, the demand for ASC diagnosis is rising, while there is a lack of specialized clinicians who are able to conduct this subjective and very time-consuming diagnosis process. 
Further, clinicians may find additional tools overwhelming, introducing the possibility of cognitive overload or misinterpretation.
We derived five design strategies that guide the development of our AI-based CDSS based on the clinicians' needs to support their decision-making and to sensitize and guide them.

\begin{figure}[!htbp]
    \centering
    \includegraphics[width=1\linewidth]{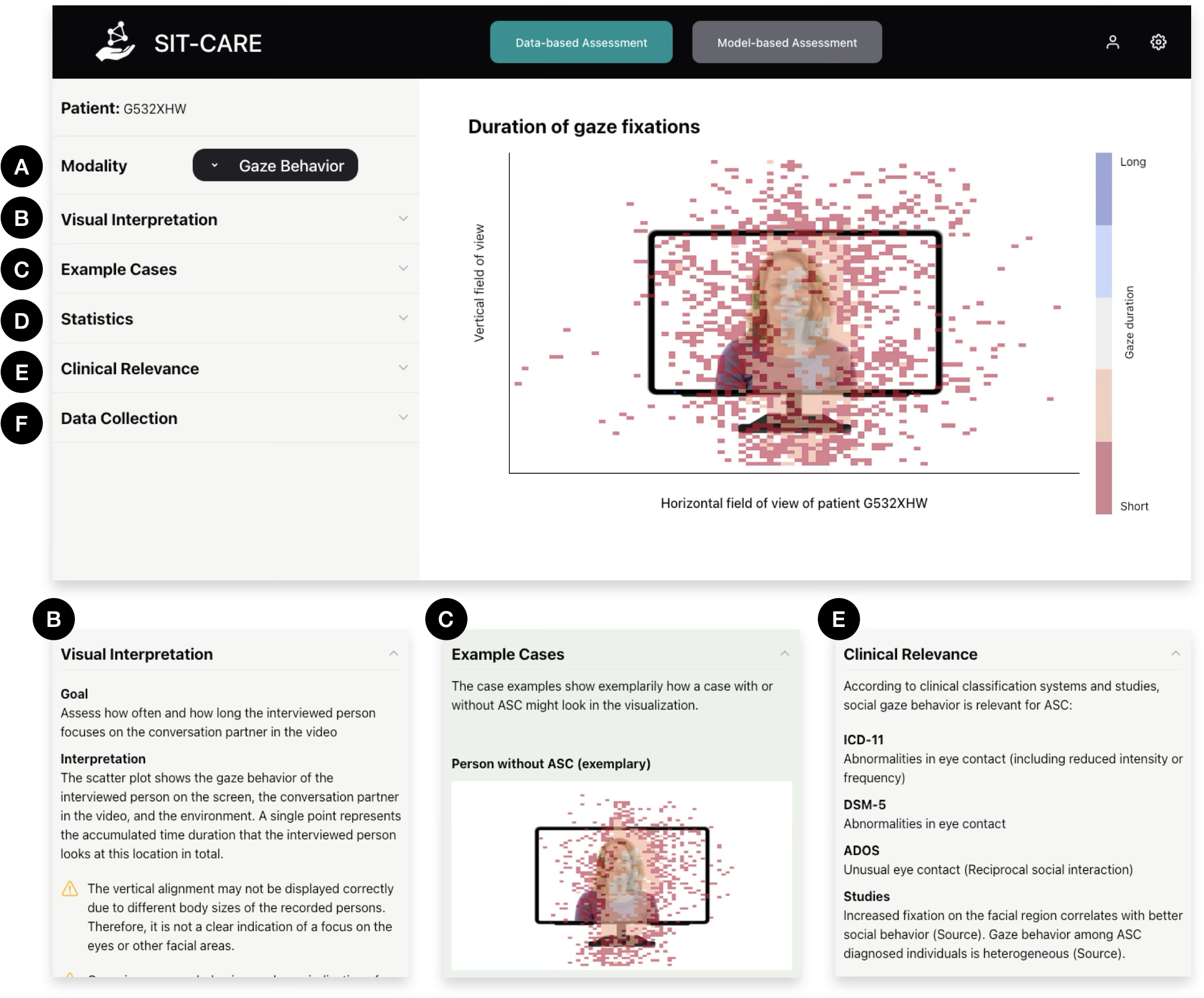}
    \caption{View of \textsc{data-based assessment} of SIT-CARE for selected modality gaze behavior. Visualization shows the gaze behavior of a hypothetical patient in relation to the screen and the actress (Image credit: \cite{drimallaAutomaticDetectionSocial2020}) during the recording in a bi-dimensional histogram plot. Left: Sidebar with collapsible additional information for the visualization and the option to change modality. Bottom: Selected additional information of the sidebar.}
    \Description[<Description of the data-based assessment mode of SIT-CARE.>]
    {<The first screenshot shows the main screen of SIT-CARE in the data-based assessment mode. This consists of a small header with navigation links to both modes of SIT-CARE, a main content area showing gaze fixation results for one hypothetical patient, and a left sidebar.  The gaze fixation results are presented as a bi-dimensional histogram plot in a coordinate system with the x-axis showing the horizontal field of view of the hypothetical patient and the y-axis showing the vertical field of view. To inform about the gaze duration, the heatmap ranges from red representing short duration to blue representing long duration. To assess where the patient looked in the middle of the coordinate system the position of the display and actress of the pre-recorded SIT video is positioned. In this case, the hypothetical patient's gaze duration is longest on the face and torso, shorter elsewhere. The left sidebar includes available modalities such as gaze behavior (A), expandable sections for visual interpretation (B), example cases (C), clinical relevance (D), statistics (E) and data collection (F). Separate screenshots show expanded sections: Visual interpretation (B): explains goal and interpretation of the visualization. Example cases (C): shows one example visualization for a person with and one without ASC. Clinical Relevance (E): explains why gaze fixation is relevant in a medical context.>}
    \label{fig:frontentd_datenbasiert}
\end{figure}

\section{SIT-CARE: Human-centered AI-based clinical decision support system for ASC screening}
\label{sec:implementation}

This section introduces the SIT-CARE system. First, we describe how the elements of the front-end user interface were developed based on the design strategies, see \autoref{sec:study1}. Then, we go through the data processing and modeling needed for both SIT-CARE modes, \ie \textsc{data-based assessment} and \textsc{model-based assessment}. Lastly, we describe the implementation of the system. In the later \autoref{sec:evaluation_study}, we evaluated SIT-CARE.

\subsection{Front-End user interface design}
\label{subsubsec:frontend_implementation}

The front-end user interface of the SIT-CARE system was developed through an iterative process. Based on the design strategies, we first created a paper prototype, adjusted the visualizations based on the clinicians feedback in the formative study, and finally translated the prototype into a functional system.
All stages were iteratively discussed among the authors, who have a background in HCI, psychology, computer science and medicine. 

\subsubsection{Structure of the user interface}

\begin{figure}[!htbp]
    \centering
    \includegraphics[width=0.8\linewidth]{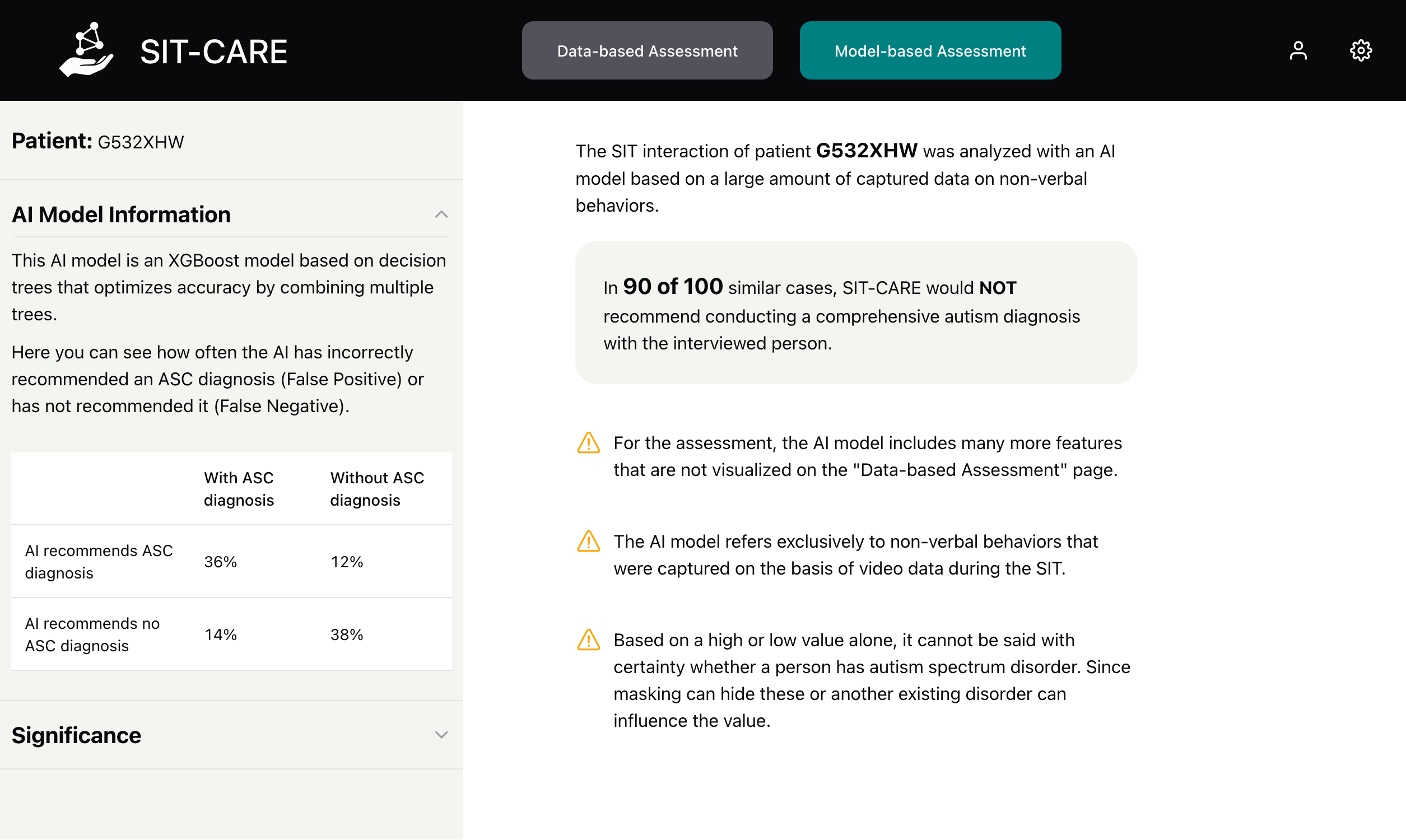}
    \caption{View of \textsc{model-based assessment} of SIT-CARE including the AI recommendation for a hypothetical patient and warnings about limitations. Left: Sidebar with collapsible additional information about the AI model and significance.}
    \Description[<Description of the data-based assessment mode of SIT-CARE.>]
    {<The screenshot shows the main screen of SIT-CARE in the model-based assessment mode. This consists of a small header with navigation links to both modes of SIT-CARE, a main content area on the right showing AI's recommendation with notes on training data and warnings that results are limited and not a definitive autism diagnosis, and a left sidebar listing hypothetical patient ID with collapsible additional information about the AI model and significance. Here, the information about the AI is expanded, explaining that an XGBoost model is used and providing a confusion matrix.>}
    \label{fig:backend_modelbasiert}
\end{figure}

As the findings of the formative study show, see \autoref{subsec:study1_findings}, clinicians do not only want an AI recommendation, but rather want to be supported in their decision-making process. As a result, SIT-CARE consists of two main modes. First, a \textsc{data-based assessment} supports clinicians in earlier decision-making stages~\cite{sox2024medical, zhangRethinkingHumanAICollaboration2024, wuCardioAIMultimodalAIbased2025} by prompting clinicians to consider ASC as part of their screening process and educating them by providing visualizations and information about nonverbal behavior, see \autoref{fig:frontentd_datenbasiert}. 
Second, the \textsc{model-based assessment} provides AI recommendations and confidence levels for the assessment of nonverbal behavior, which aids in determining whether full ASC diagnostics is needed,  \autoref{fig:backend_modelbasiert}.
As a result, SIT-CARE is intended to support gathering data to explore and weigh whether nonverbal behavior is typical for ASC or not, and provides a direct suggestions to guide ASC screening (\textit{Design Strategy 1}). 
%
Additionally, a start screen clarifies that only nonverbal behavior is evaluated and that the system is not intended for standalone usage, aligning with \textit{Design Strategy 5} and the human-AI interaction guideline ``Make clear what the system can do''~\cite{amershiGuidelinesHumanAIInteraction2019}.
Furthermore, to avoid information overload \textit{(Design Strategy 3)}, additional system information in the sidebar is progressively disclosed  by collapsing all categories in a drop-down list by default~\cite{springerProgressiveDisclosureWhen2020}, which can lead to higher user acceptance in time-constrained environments~\cite{jacobsDesigningAITrust2021, wangHumancenteredDesignEvaluation2023}.
In the following, both assessment modes of the system are described.

\subsubsection{Data-based assessment mode}

The \textsc{data-based assessment} mode of SIT-CARE allows clinicians to explore nonverbal behavioral data of a patient that was derived from the SIT video. 
Following the advice of clinicians, we only considered modalities in which they expressed interest and that are also relevant in current classification systems (ICD-10~\cite{Organization2004_ICD-10}, ICD-11~\cite{ICD11}, DSM-5~\cite{alma9918660570001341}, ADOS~\cite{lordAutismDiagnosticObservation}) and research (\eg \cite{fusaroliVoiceMarkerAutism2017, riddifordGazeSocialFunctioning2022, trevisanFacialExpressionProduction2018}). This led to the inclusion of eye gaze behavior~\cite{riddifordGazeSocialFunctioning2022}, intensity and variability of facial expressions~\cite{trevisanFacialExpressionProduction2018}, and voice parameters~\cite{fusaroliVoiceMarkerAutism2017} \textit{(Design Strategy 3)}. 
Providing such insights into the training data may increase transparency, which could improve clinicians' understanding of the AI model in the \textsc{model-based assessment}~\cite{laiScienceHumanAIDecision2023a}.

For a given patient, the main content in \textsc{data-based assessment} mode shows a visualization of the currently selected modality, \ie gaze behavior, facial expression or voice (\autoref{fig:frontentd_datenbasiert} A).
See \autoref{fig:frontentd_datenbasiert} for gaze behavior, where a bi-dimensional histogram plot is used. This visualization was adjusted based on clinicians feedback to include an illustration of the screen and the position of the actress\footnote{The image of the actress shown is from~\cite{drimallaAutomaticDetectionSocial2020} published under the CC BY 4.0 International License, and has been adjusted by us (background color change, cropped).} in the SIT, to make the interpretation more intuitive \textit{(Design Strategy 2)}. 
Line plots were used to depict the variability and intensity of facial expressions over time, which were improved through clinicians' suggestions to include a reference area based on standard deviations from the reference group mean to make it more understandable \textit{(Design Strategy 2)}, see \autoref{sec:app_visualizations}.
For voice parameters, we focused on pitch variability.
Based on clinician feedback, initial box plots were replaced with violin plots to visualize pitch variability with respect to Non-ASC reference distributions \textit{(Design Strategy 2)}, see \autoref{sec:app_visualizations}.
The line and violin plots were divided into three sections (neutral, joy, disgust) corresponding to the interaction phases of the SIT.
Via radio buttons, clinicians could then compare the line and violin plots of the current patient case to different, gender-specific reference groups (\textit{Design Strategy 4}).
A hover element provides additional information about the reference groups (\textit{Design Strategy 1}). 

To further guide clinicians, a sidebar with additional information on the current modality is integrated; see \autoref{fig:frontentd_datenbasiert} on the left. 
Under ``Visual Interpretation``, see \autoref{fig:frontentd_datenbasiert} B, information is provided on how to read the current visualization in order to understand and interpret it correctly (\textit{Design Strategy 1}).
Following \textit{Design Strategy 4}, prototypical examples~\cite{poche2023natural} of visualizations for an individual with or without ASC are provided, which allows clinicians to compare exemplary case data with the current patient (\autoref{fig:frontentd_datenbasiert} C). 
%
Summary statistics of the patient for the current modality are presented in a tabular format alongside group-level means and standard deviations of ASC and Non-ASC reference groups (\autoref{fig:frontentd_datenbasiert} D). 
To connect to familiar concepts and guide less experienced clinicians (\textit{Design Strategy 1 and 4}), we provide information about the clinical relevance per modality (\autoref{fig:frontentd_datenbasiert} E).

\subsubsection{Model-based assessment mode}

Switching to the \textsc{model-based assessment} mode allows clinicians to inspect the output of the binary AI classification model, which provides a recommendation whether a comprehensive diagnostic procedure for ASC should be conducted \textit{(Design Strategy 1)}. 
Limitations were communicated via warnings below the recommendation. For example, clinicians were reminded that additional features other than the ones presented in the \textsc{data-based assessment} were used and that masking should be considered \textit{(Design Strategy 5)}.
To improve interpretability, the recommendation is presented in frequency format~\cite{cao2024designing}, indicating the confidence in the form of a predicted probability of the AI model and therefore revealing ``information about the prediction''~\cite{laiScienceHumanAIDecision2023a}. 
``Information about the model''~\cite{laiScienceHumanAIDecision2023a} is provided via the left sidebar, such as insights into the accuracy of the model with a contextualized confusion matrix~\cite{shenDesigningAlternativeRepresentations2020a, amershiGuidelinesHumanAIInteraction2019} (``AI Model Information''). Further, under (``Significance''), information about how the model's accuracy changes if, for example, only gaze behaviors would be considered is provided \textit{(Design Strategy 5)}.

\subsection{Data processing and modeling}
\label{subsec:implementation_backend}

During the SIT, the patient is video- and audio-recorded. From these recordings, nonverbal behavioral features were extracted.
All extracted features were aggregated into phase-specific representations (listening vs. speaking) across emotion-eliciting interaction phases (neutral, joy, disgust), enabling targeted analysis of both expressive and receptive behavior. 
In SIT-CARE, these extracted features are used to generate visualizations and numerical summaries of gaze behavior, facial expressions, and voice parameters within the \textsc{data-based assessment}. 
Further, for \textsc{model-based assessment}, we trained AI models using additionally derived nonverbal features for ASC classification beyond the three introduced modalities.
%
In the next section, the back-end is described. 
Detailed formulas and sampling procedures are provided in \autoref{app:calculations}.

\subsubsection*{Gaze behavior:}
Eye gaze behavior was extracted using OpenFace 2.2~\cite{Baltrusaitis2018} gaze angle estimates and then geometrically transformed into a screen-centered coordinate system, enabling the computation of metrics such as gaze variability, fixation duration on the display or actress in the video, and frequency of gaze aversion. 
New transformed gaze angles were also used to visualize fixation locations relative to the actress as bi-dimensional histogram plots.
Bi-dimensional histogram plots were generated by color-coding fixation density into five discrete levels, defined relative to a Non-ASC reference group. 
Prototypical examples from individuals of the ASC and the Non-ASC groups were sampled. In addition, the variance of horizontal gaze angles and the percentage of time spent looking at the screen and at the actress’s face region were calculated. For details see \autoref{app:calculations}.

\subsubsection*{Facial expressions:}
Facial expressions were extracted using OpenFace 2.2~\cite{Baltrusaitis2018} and quantified via facial action units (AUs). We focused on facial expression representing smiling intensity (AU06 cheek raiser, AU12 lip corner puller).
For each assessed individual, the mean AU intensity traces were plotted across the interaction using line plots. To allow comparing the current case with Non-ASC individuals, two reference bands were overlaid for comparison: a dark band representing one, and a lighter band representing two standard deviations. 
Representative individuals were selected using a distance-to-median score and visualized as prototypical examples, with 15\% noise added to the traces to ensure anonymity. Further, the mean and variance of smiling intensity (AU06 + AU12), the mean and variance of AU04 (brow lowering), and the overall mean intensity across all AUs were calculated. For technical details see \autoref{app:calculations}.

\subsubsection*{Voice parameters:}
Prosodic aspects of speech were extracted using the openSMILE~\cite{Eyben2010} toolkit, including the features pitch, intensity, jitter, shimmer, and harmonic-to-noise ratio. 
Pitch variability measures the range of fundamental frequency (F0, in semitones relative to 27.5 Hz) between the 2nd and 98th percentiles of voiced segments. 
Prototypical example cases corresponded to group medians.
Further, pitch variance, mean, and variance of voiced segment length, and loudness variability were calculated. For more information, see \autoref{app:calculations}.

\subsubsection*{AI classification model:}
Beyond the selected features used in the \textsc{data-based assessment} mode, additional multimodal features were used to train a binary classification model (ASC, Non-ASC). 
For instance, head motion parameters were additionally extracted and processed using the OpenFace library. 
In total, 1,140 features were used to train a late-fusion binary classification model. 
Thus, each modality was modeled separately using XGBoost (gradient-boosted decision trees). The resulting probability scores were combined using logistic regression with polynomial features (degree 2) to capture non-linear interactions. 
We trained the model on a dataset of 325 adult participants (168 ASC, 157 Non-ASC; gender-balanced), which was also used for the reference group information above. This large-scale video dataset was collected with IRB approval. Inclusion criteria encompassed age (18-65), IQ over 85, language fluency, no or stable pharmacotherapy and no psychiatric comorbidities.
The model was then applied to classify two patient cases with their consent that were not part of the training dataset, which the model classified correctly. 
Performance was evaluated using participant-based Leave-One-Out Cross-Validation~\cite{hastie2009elements} to ensure robust generalization.  
In that evaluation, the late-fusion model achieved an accuracy of 74\%, with a precision of 0.76 and a recall of 0.74, outperforming unimodal approaches.

\subsection{System implementation}
\label{subsec:system_implementation}

The SIT-CARE system was developed as a typescript-based web application. For the front- and back-end implementation, the NextJS/React\footnote{https://nextjs.org/} framework was used, ChakraUI\footnote{https://www.chakra-ui.com/} was selected as visual component library. 
Visualizations of patient modalities were generated in Python using matplotlib\footnote{https://matplotlib.org/} and seaborn\footnote{https://seaborn.pydata.org/} and then integrated into the web application. While contextual elements such as legends, layouts, and interactions with visualizations were implemented directly in the web application, the visualizations needed to be pre-rendered due to the high data privacy requirements of working with actual ASC patient data. 
The application was deployed on institutional servers, and study participants were granted access through password-protected links.

\section{Evaluation study: Understanding and supporting AI-assisted clinical decision-making}
\label{sec:evaluation_study}

To investigate the influence of SIT-CARE on clinicians' mental models and reasoning during their decision-making and to answer our second research question, we conducted a study with clinicians during which they interacted with SIT-CARE and assessed two patient cases. 
Per patient case, at three decision points (DP), they made a decision each: after seeing the SIT video (DP1), after exploring SIT-CARE's \textsc{data-based assessment} (DP2), and lastly, after receiving the \textsc{model-based assessment} (DP3).
Our findings summarize what decision paths clinicians took and how their mental models of the AI and reasoning played a role in each decision. 
Further, clinicians expressed high interest in using SIT-CARE in their ASC screening or as a learning opportunity.

\subsection{Method}

\begin{table*}[t]
    \captionsetup{justification=centering} 
    \small{
    \caption{Demographics of Participants in our Evaluation Study.}
    \Description{This table presents the demographics of participants in our evaluation study. The table has 4 columns: P#: A unique identifier for each participant. Gender: The gender of each participant. Professional Stage: The stage of professional development of each participant, which can be either a psychotherapist in training or a practicing psychotherapist. Experience: The amount of experience each participant has in their profession, including the number of years and whether they are specialized in ASC. The table has 7 rows, each representing a different participant, P8 to P14. Here are the values listed per row P#: P8 is a female practicing psychotherapist with 4 years of experience. P9 is a female practicing psychotherapist with 40 years of experience, with experience working with ASC. P10 is a male practicing psychotherapist with 12 years of experience. P11 is a male practicing psychotherapist with 30 years of experience, specialized in ASC. P12 is a male practicing psychotherapist with 40 years of experience, specialized in ASC. P13 is a female practicing psychotherapist with 11 years of experience, specialized in ASC. P14 is a female practicing psychotherapist with more than 4 years of experience, with first experiences working with ASC.
    }
    \label{tab:demographics_2}
    \renewcommand{\arraystretch}{1.5}
    \begin{tabular}{ccp{6cm}p{5cm}}
        \Xhline{3\arrayrulewidth}
        \textbf{P\#} & 
        \textbf{Gender} & 
        \textbf{Professional Stage} & 
        \textbf{Experience}
        \\ 
        \Xhline{2\arrayrulewidth}
        P8 & 
        Female & 
        Practicing psychotherapist  & 
        4 years
        \\ 
        P9 & 
        Female & 
        Practicing psychotherapist  & 
        40 years, experience with ASC
        \\
        P10 & 
        Male & 
        Practicing psychotherapist  &  
        12 years
        \\
        P11 & 
        Male & 
        Practicing psychotherapist  & 
        30 years, specialized in ASC
        \\
        P12 & 
        Male & 
        Practicing psychotherapist  & 
        40 years, specialized in ASC
        \\
        P13 & 
        Female & 
        Practicing psychotherapist  & 
        11 years, specialized in ASC
        \\
        P14 & 
        Female & 
        Practicing psychotherapist  & 
        >4 years, first experiences with ASC
        \\
        \Xhline{3\arrayrulewidth}
    \end{tabular}%
    }
\end{table*}

We newly recruited seven clinicians with differing experience in clinical psychology and ASC via purposive sampling~\cite{andradeInconvenientTruthConvenience2021}, see \autoref{tab:demographics_2}.
We conducted open-ended, semi-structured interviews~\cite{longhurst2003semi} remotely with two interviewers, each lasting around 70 minutes.
The study was approved by our IRB (application number, institution, and date will be provided after acceptance). All participants consented prior to participation.

During the interviews, we first asked demographic and work-related questions, then introduced SIT-CARE.
Next, the clinicians were asked to assess two patient cases, which were chosen by experienced clinicians based on their diagnostic assessments (first case: Non-ASC, second case: ASC), and are considered clear cases. We obtained the patients' informed consent to share their data.
For each case, starting with the Non-ASC case, the participants first saw the video recording of the individual conducting the SIT. 
Next, they explored SIT-CARE's \textsc{data-based assessment} and then the \textsc{model-based assessment} (see \autoref{sec:implementation}).
After each step, the clinicians were asked if the patient's nonverbal behavior indicated ASC and to explain their reasoning.
Thus, per patient case, the clinicians made a decision at each decision points (DP): after seeing the SIT video \textbf{(DP1)}, after exploring SIT-CARE's \textsc{data-based assessment} \textbf{(DP2)}, and after receiving the \textsc{model-based assessment} \textbf{(DP3)}.
To explore their mental models about the AI, participants were asked to predict what the AI might suggest and why at DP1 and DP2 (see \cite{hoffmanMeasuresExplainableAI2023}).
During the task, participants verbalized their cognitive process (think-aloud protocols), which has been shown to be useful for exploring decision processes while interacting with CDSS~\cite{toddProcessTracingMethods1987, abdel-karimHowAIBasedSystems2023, vitalariKnowledgeBasisExpertise1985}.
Additional questions were asked, such as about the helpfulness of SIT-CARE and their general impression, see \autoref{sec:app_interviewscript_2}.

Approximately eight hours of audio material were transcribed, which we inductively and deductively coded~\cite{mayring2014content}.
We analyzed the transcripts by reading them multiple times and derived the initial codes iteratively.
We coded the materials for each decision point and patient case. The decisions could either be ASC, Non-ASC, or unsure, thus, could be directly derived from the transcripts (deductive).
To understand the participants' decision-making process, we grouped the decision paths that the clinicians followed based on their decisions made at each decision point. This allowed us to explore the mental models that led a clinician to take a specific path. Clinicians' feedback on SIT-CARE modes and purpose was coded inductively.
The findings were collaboratively structured and refined through iterative feedback and team discussions to ensure consensus.

\subsection{Findings: Evaluation and integration of the SIT-CARE system}

In this section, we summarize the clinicians' opinions about SIT-CARE, \ie the \textsc{data-based assessment}, followed by the feedback about the \textsc{model-based assessment}, and lastly, the integration and use of SIT-CARE in practice.

\subsubsection{Diverse expectations of data-based assessments}
\label{subsubsec:study2_findings_data}

Overall, most participants found the \textsc{data-based assessment} to be helpful and reliable. Especially the gaze visualization and the prototypical examples were liked. However, opinions on the level of detail varied.

The gaze information was described as intuitive and ``\textit{really good represented graphically}'' (P8).
However, the violin plots were described as rather unfamiliar (P8, P14), and the reference area for the line plots could be misunderstood (P11).
Participants had different opinions about the type and level of detail of the information.
Some participants expressed interest in detailed summary statistics (\eg per SIT phase and gender-specific (P12)), analysis of gestures and what was said (P8, P11). 
In contrast, others expressed that the different types of visualizations may already be challenging. P10 stated: ``\textit{The less familiar the person is with scientific work, the more difficult it will probably be to understand such graphs}''. 
P13 recommended to focus only on ``\textit{one [nonverbal behavior] that is classically presentable, for example, gaze behavior}'' and underlined that detailed statistics would not be helpful.
%
In addition, participants mentioned that they need an ``\textit{introductory module to sort out this mass of information and data [...] so that [they] know what to do with all the information}'' (P13) and more information on ``\textit{why [some behavior] is somehow typical for autism}'' (P14). 
To guide clinicians, P10 recommended highlighting in the visualizations ``\textit{what you would expect to stand out from a normal, [Non-ASC] control group [...]}''.
In addition, the prototypical examples were received very well, and it was recommended to include more examples depicting a range of possible ASC behavior (P10, P11).

\subsubsection{Feedback on the model-based assessment}
\label{subsubsec:study2_findings_model}

Participants had mixed opinions on the \textsc{model-based assessment}, with some disregarding it completely and others considering it more valuable than the \textsc{data-based assessment}.

Two participants reported that they ``\textit{ignored}'' (P9) the \textsc{model-based assessment} and ``\textit{found [one case] to be misjudged}'' (P11).
Others emphasized the advantage of \textsc{the model-based assessment}, which is ``\textit{time-efficient}'' (P13) and provides ``\textit{additional information that the normal human brain cannot fully process}'' (P13) without help,  ``\textit{while the data-based assessment only exemplarily represents three aspects}'' (P10). 
P12 described that the \textsc{model-based assessment} interrupted their typical decision-making process: ``\textit{I was deviating from my usual assessment, and I felt uncertain. [...] There's the value, and what does that actually mean, and does that align with my impression or the questionnaire values? So that irritated me a little at first, which isn't such a bad thing, to recalibrate yourself in diagnostics as well. It creates a bit of awareness to take a closer look at yourself.}''    
%
Some participants preferred such ``\textit{a clear value}'' (P13), others warned that a number as ``\textit{bit tempting}'' which may lead to over-reliance, ignoring aspects like masking (P12).
This risk was partially mitigated by integrating a confidence matrix and warnings, which were noted as ``\textit{very transparent, good, and important}'' (P10). 
Further adjustments were recommended, such as providing more information about the AI model (P9), considering gender more (P12), and what was said (P11).
Participants also discussed whether they want to be able to calibrate the AI model's sensitivity and specificity. However, P13 argued that doing so would hinder standardization.

\subsubsection{Between supporting ASC screening and raising awareness}

Overall, participants expressed high interest in SIT-CARE as an ASC screening support and as a learning opportunity.

All participants expressed interest in SIT-CARE and indicated that they would use it ``\textit{Yes, immediately. Yes, absolutely.}'' (P12), and were open for AI-based support ``\textit{if there's a trustworthy organization behind it}'' (P12).
%
Two purposes of SIT-CARE were highlighted: 
First, to use it in everyday practice to support ASC diagnostics and screening as the ``\textit{autism screening instruments that currently exist [...] are simply so outdated that they don't give much anymore}'' (P13), and second, to raise awareness and educate oneself about the topic (P10, P12, P13). 
P12 stated ``\textit{that it can definitely help raise awareness of a possible ASC diagnosis that I would otherwise overlook}''.
Further, all participants expressed a need for guidance in the form of additional information and an onboarding process. 
For example, P13 suggested ``\textit{a short video would be helpful to explain [the data-based assessment]}'' (P13).
%
Moreover, in both studies, participants expressed an interest in expanding SIT-CARE beyond ASC.
P2 were unsure if it is possible to differentiate between disorders, but ``\textit{[they] think there are probably specific characteristics that apply to specific disorders}'' and that ``\textit{whether that's really specific would be totally cool to find out.}''

\begin{figure}
    \centering
    \includegraphics[width=0.90\linewidth]{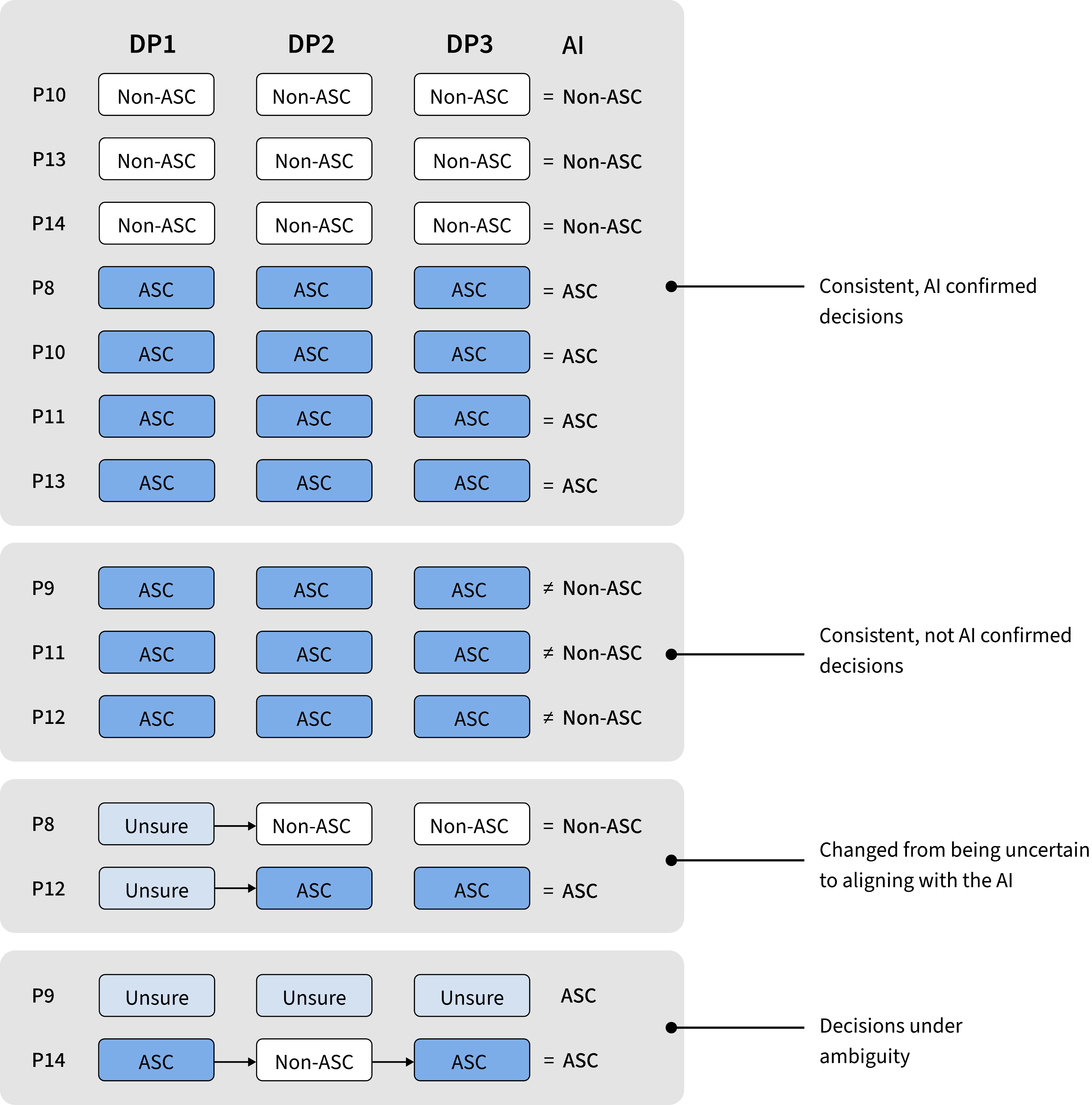}
    \caption{Depiction of the four groups of decision paths the newly recruited seven clinicians took per case. Each clinician assessed two cases, leading to 14 single decision paths and 42 single decisions.
    Per decision path the participant number (P8-P14), their three decisions per case (DP1, DP2, DP3) and the AI recommendation (on the right) is provided. 
    Decision changes of the clinicians over the three decision points, \ie video-based assessment (DP1), data-based assessment (DP2), and model-based assessment (DP3) are marked with arrows from one decision point to the subsequent decision point.}
    \label{fig:study2_decisionsv2}
    \Description[<Depiction of the decision paths and groups described in the findings of the evaluation study.>]{<This diagram illustrates the decision paths of seven clinicians. Each clinician assessed two cases, one ASC and one without Non-ASC. In total 14 decision paths, each consisting of three decisions (DP1, DP2, DP3) are shown. The decisions at each decision point are either ASC, Non-ASC or unsure. Per decision path the participants making the decisions and the AI recommendation, which aligns with the ground truth is stated. The decision paths are grouped in four groups, i.e., (1) consistent, AI confirmed decisions, (2) consistent, not AI confirmed decisions, (3) changed from being uncertain to aligning with the AI, and (4) decisions under ambiguity. In the first decision path group, P10, P13 and P14 decided at DP1, DP2 and DP3 on Non-ASC, which aligned with the AI. P8, P10, P11 and P13 decided at DP1, DP2 and DP3 on ASC, which aligned with the AI. In the second decision path group, P9, P11 and P12 decided at DP1, DP2 and DP3 on ASC, which did not align with the AI. In the third decision path group, P8 decided at DP1 on unsure and at DP2 and DP3 on Non-ASC, which did not align with the AI. P12 decided at DP1 on Unsure and at DP2 and DP3 on ASC, which did not align with the AI. In the fourth decision path group, P9 decided at DP1, DP2 and DP3 on Unsure, while the AI recommended ASC. P14 decided at DP1 on ASC, at DP2 on Non-ASC and at DP3 again on ASC, the AI recommended ASC. 
    >}
\end{figure}

\subsection{Findings: Decision-making process and users' mental models}
\label{subsec:study2_findings_decisionprocess}

To answer our second research question (RQ2), in the following, we describe the decision paths the clinicians followed during their decision-making process per patient case. 
Seven clinicians assessed each of the two patient cases at three decision points, thus making 42 decisions. 
In each assessment of the patient case, the clinicians took a specific path, which is characterized by what decision was made at each decision point, and whether the decision was changed after interacting with SIT-CARE, see \autoref{fig:study2_decisionsv2}.
We summarize these decision paths and explore clinicians' mental models during this process.

\subsubsection{Different reasoning can lead to the same decision}
\label{subsubsec:study2_findings_agreement}
%
In seven of the fourteen decisions to be made, the clinicians did not deviate from their initial decision, which was then confirmed in the \textsc{model-based assessment}.
Thus, the participants made \textbf{consistent, AI confirmed decisions}.
While their decision paths were similar, they considered SIT-CARE information differently.

P14 interacted extensively with the \textsc{data-based assessment}, read all materials in detail, considered the prototypical examples and summary statistics: ``\textit{I find the data helpful because it captures much more than I would notice in person}''.
In contrast, P13 only considered their own and the \textsc{model-based assessment} in both of their assessments and stated ``\textit{the way the AI collects its data is so different from my way of perceiving this information, so far apart, that I wouldn't even try to compare it.}'', thus they did not see the need to build a sound mental model of the AI system.
Regarding the second case (ASC), only when following this consistent, AI confirmed decision path, participants highlighted the person's ``\textit{stimming behavior}'' (P11), ``\textit{odd hand movement}'' (P13) and ``\textit{stereotypical movements}'' (P8), \ie ASC typical movements that are currently not part of our AI system.
For all of the described decisions above, the participants predicted the AI to come to the same conclusion and confirm their assessment.
While P8's prediction of the AI was correct, they had an incorrect mental model of the AI as they optimistically predicted a high AI confidence, and were surprised by the real confidence value, stating they are ``\textit{not so familiar with the numbers}''.

\subsubsection{Perceived conflict between the data-based and model-based assessment}
\label{subsubsec:study2_findings_disagreement}

Three participants rated the first patient case (Non-ASC) as rather typical for ASC at each decision point, disagreeing with the \textsc{model-based assessment}.
Thus, they made \textbf{consistent, not AI confirmed decisions}. Although the \textsc{data-based assessment} prompted reflection, the video-based impressions were emphasized more.

After seeing the video (DP1), they considered not only nonverbal behavior, but also what was said. 
P9 specified: ``\textit{I don't think others [without ASC] would quickly grasp, the neutral flavor that has been extracted [.. ], I found that to be typically autistic.}''
After reviewing the data-based assessment, one participant was still confident in their decision (P11), others mentioned ``\textit{a few aspects, like this gaze into the eyes, [as] rather untypical for autism}'' (P12).
Despite identifying such contraindicating cues in the \textsc{data-based assessment} that provided partial insight into the training data based on their mental model of the AI, they still assumed that SIT-CARE would confirm the contraindications. They were surprised when it did not.
P11 stated ``\textit{that irritates [me] because [...] that was relatively autism-typical.}''.
P9 was still unsure about the AI's reasoning, but hypothesized that: ``\textit{It sounds like the AI is looking at the same thing, but maybe with a different threshold or weighting''} (P9) as they would consider the case as a ``\textit{light expression of autistic characteristics}''.
All three participants considered the \textsc{data-based assessment} (P9, P12) and their video-based assessment as most insightful, P9 stated that they ``\textit{trust [themselves] the most}'' (P9).

\subsubsection{From indecision to decision}

In two decision paths, participants \textbf{changed from being uncertain to aligning with the AI} after exploring the \textsc{data-based assessment}, which helped them find indicators that led to a more confident assessment.

Based on the video for the second patient case (ASC), P12 stated: ``\textit{I didn't find it clearly typical [...] as there was too much there.}''. Another participant noticed contradicting indications, as the person smiled appropriately but also seemed ``\textit{petrified}'' (P8).
Both were unsure of what the AI may recommend but speculated in detail about what the AI would recommend depending on whether the AI considers what was said (P8, P12), masking, or gender (P12). 
Further, P8 mentioned a strength of the AI: ``\textit{I think what it could do even better is to filter out facial expressions even better and detect smaller movements that we can't perceive}''.
After exploring the \textsc{data-based assessment}, their decisions changed.
P8 stated that ``\textit{It kind of confirms me, because I was totally unsure}'' and now assumed it is ``\textit{a person without autism, because they actually show a lot more in their facial expression or even matching facial expressions to what's happening.}'', but refrained from predicting the AI's recommendation. 
P12 said the \textsc{data-based assessment} clarified the position of the display and that ``\textit{with the individual data, the scales tipped slightly in the direction of autism, so I would have recommended further diagnostics.}''.
The \textsc{model-based assessment} manifested their updated decisions. 
P12 reiterated the data visualization of the eye behavior to be ``\textit{a key that strengthened my tendencies}''.

\subsubsection{Decisions under ambiguity}

Two decision paths were shaped by \textbf{ambiguity}, showing the risk of indecision and the lack of comprehensive comparisons to other diagnoses.

For the second case (ASC), P9 was indecisive at all three decision points and, after seeing the video (DP1), explained that the person appeared ``\textit{not particularly lively, but that can also be due to depression}'', reiterated their surprise about the AI's first recommendation, and that they can not predict the AI. 
After seeing the \textsc{model-based assessment} (DP3), they underlined: ``\textit{For me, depression is simply in the foreground}''.
In contrast, P14 changed their decision twice, first leaning towards ASC based on gaze and facial expressions (DP1), then changing their mind (DP2) because ``\textit{if you look at the numbers, it will be more untypical [for ASC] from the facial expression and also from the voice.}'' 
At each DP, they assumed the AI would confirm their decision. After seeing the model-based assessment (DP3), they rationalized the AI recommendation, stating ``\textit{that eye contact is perhaps the most important factor [for the AI] then it makes sense [...] to carry out more detailed diagnostics}'' and changed their decision back to ASC.

\section{Discussion}

We designed SIT-CARE based on the findings of the formative study that explored clinicians’ workflows and needs. SIT-CARE presents nonverbal behavioral features from SIT videos in two modes. The \textsc{data-based assessment} provides multiple visualizations and numerical summaries of gaze behavior, facial expressions, and voice parameters. The \textsc{model-based assessment} presents an AI recommendation on whether to perform comprehensive ASC diagnostics. In an evaluation study, we examined how SIT-CARE influences clinicians' mental models and decision-making processes.

\subsection{Clinicians' decision paths and mental models}

In our evaluation study, we presented two real patient cases: one with an ASC diagnosis and one without. Seven clinicians assessed each case at three decision points, \ie after seeing the SIT video (DP1), after the \textsc{data-based assessment} (DP2), and lastly, after the \textsc{model-based assessment} (DP3).
Thus, in total, 42 decisions were made across 14 decision paths. For each decision on a decision path, we investigated the clinicians' current mental models of the AI and their reasoning. 

The first group (see \autoref{fig:study2_decisionsv2}, top group) accounts for half of all decision paths. In this group, the clinicians' initial decisions remained unchanged and were confirmed by the AI recommendation. 
Clinicians seemed to reflect on the capabilities and functioning of the AI system. 
While some recognized that the AI captured more than they would have noticed and found the \textsc{data-based assessment} helpful, others highlighted aspects that SIT-CARE did not consider and thus recognized the AI's limitations. Consequently, they developed different mental models of the AI's capabilities. 
Further, one clinician expressed no interest in building a mental model of the AI and only valued the recommendation, while others explored all information intensely and weighted the information of SIT-CARE with aspects not considered by this system. 
While their decisions aligned with their prediction of the AI recommendation, we found that clinicians did expect the AI's confidence to be more confirming. As AI confidence seems to influence human confidence~\cite{liConfidenceAlignsUnderstanding2025}, training is needed to help clinicians handle this information.
Contrary to this, another group of decision paths is defined by clinicians constantly misjudging the Non-ASC case as ASC (see \autoref{fig:study2_decisionsv2}, second-highest group).
Similarly to the first group, some clinicians noticed conflicts between their impression and the \textsc{data-based assessment}. Others speculated that the AI reasons similarly to them, but decides based on different thresholds. However, here, they did seem to consider the AI's strengths and limitations less, which may indicate a less nuanced understanding of the AI and potential under-reliance. 
Prior research has shown that educating users about the strengths and limitations of the AI and themselves, can improve human-AI collaboration~\cite{cai2019, holsteinSupportingPerceptualComplementarity2023, pinski2023}. In future research, strengthening their mental model of AI capabilities may improve their decision-making process.

We identified two further groups of decision paths (see \autoref{fig:study2_decisionsv2}, bottom two). While in the third group SIT-CARE led from indecision to decision, in the fourth group, the opposite seemed to occur.
The third group was indecisive after the video (DP1), but became more decisive after the \textsc{data-based assessment}. 
Similar to the first group, the clinicians taking this decision path seemed to acknowledge the strengths of the AI, such as detecting small details.
In the fourth group, one clinician did not consider the AI recommendation, because they had a different diagnosis in mind. Another clinician's assessment was unstable, changing their decision after receiving new information.
Summarized, while the first and third group show that SIT-CARE can support creating an appropriate mental model of the AI and consolidate clinicians' reasoning, the second and fourth decision groups indicate that the system can also lead to confusion and over- and under-reliance.

\subsection{Standardized screening support and learning opportunity}

Clinicians highlighted in the evaluation study two primary application areas for SIT-CARE: (1) supporting the diagnostic process, and (2) guiding less experienced clinicians to evaluate nonverbal behaviors associated with ASC.

To support the diagnostic process, SIT-CARE, specifically the \textsc{model-based assessment}, could be used during the screening stage, \ie before detailed diagnostics. It could help clinicians determine which individuals should be referred to specialists, replacing or complementing less reliable questionnaires (see \autoref{subsec:RW_ASC_AI_Assistance}). 
Since the SIT can be performed at home in a few minutes, SIT-CARE has strong scalability potential.
Furthermore, clinicians can use SIT-CARE regardless of their location, enabling its future integration into telehealth workflows.

The guidance and training potential of SIT-CARE is highly relevant given common misconceptions of ASC among general practitioners \cite{wrightShouldAutismSpectrum2020} and the shortage of specialized clinicians~\cite{lipinskiBlindSpotMental2022, volkmar2022diagnostic}. 
By enabling the exploration of gaze behavior, facial expressions, and voice parameters, as well as by comparing the current patient case with prototypical cases, SIT-CARE's \textsc{data-based assessment} could help less experienced clinicians develop diagnostic sensitivity for assessing nonverbal behaviors of patients. To achieve this, a training phase or tutorials can be included to educate clinicians~\cite{morana2017, lai2022a}.
From a clinical perspective, it is notable that SIT-CARE enables clinicians to compare the current patient case with gender-specific reference groups. This raises awareness of the female autism phenotype, \ie that autism is expressed differently in females and is therefore often overlooked~\cite{bargiela2016experiences, volkmar2022diagnostic, hull2020female}.

\subsection{Ethical considerations in using AI for the diagnosis process of ASC and beyond}

Our research focused on creating an AI-based CDSS for ASC assessment based on the SIT. However, using such a tool in a clinical setting raises a number of ethical concerns.
Due to long waitlists and workforce shortages, institutions may use AI outputs as de facto diagnoses, which contradicts the intended role of SIT-CARE as a decision support tool. Therefore, it is important to position SIT-CARE as a complement to, rather than a replacement for, comprehensive ASC assessment, for example, through additional scope-of-use messaging (``recommendation to consider further diagnostics'' rather than ``diagnosis''~\cite{Song2019_TheUseofArtificialIntelligenceinScreeningandDiagnosisofAutismSpectrumDisorder}).
In addition, clear clinical AI usage guidelines are needed~\cite{kolbingerReportingGuidelinesMedical2024a}. For example, clinicians could be required to document how they considered SIT-CARE's \textsc{data-} or \textsc{model-based assessment} in their decision-making process.
Furthermore, incorporating supervised onboarding and periodic training is needed to allow appropriate usage, reflection, and criticality by clinicians~\cite{morana2017}.
Our study also revealed that clinicians expressed that the trustworthiness of the institution behind SIT-CARE matters to them. Patient data is sensitive, so a CCPA\footnote{The California Consumer Privacy Act (CCPA) is a United States law that provides California residents with greater control over their personal data. For more information, please visit \url{https://oag.ca.gov/privacy/ccpa}.} or GDPR\footnote{The General Data Protection Regulation (GDPR) is a European Union law that governs how personal data is collected, processed, and stored, emphasizing transparency, security, and individual rights over their data. For more information, please use \url{https://gdpr-info.eu/}.}-compliant data security policy is indispensable to protect against unauthorized access and data breaches.

\section{Limitations and future work}

Our work has several limitations.
%
First, we involved seven clinicians in our formative study and newly recruited seven clinicians in our evaluation study. To receive professional feedback, most of our recruited clinicians have several years of experience with ASC.
Even though our sample size is similar to prior domain-specific qualitative work (\eg \cite{cai2019, zhangRethinkingHumanAICollaboration2024, wuCardioAIMultimodalAIbased2025, beedeHumanCenteredEvaluationDeep2020a}), future work is needed to specifically investigate the generalizability to other target groups, such as inexperienced clinicians.

Second, interviewing clinicians provided valuable insights into their decision-making processes and mental models. However, large-scale quantitative research with clinicians via cluster sampling is needed to assess team performance and to explore the decision paths under real-life circumstances. 
Thus, our research provided the foundation, but future work requires a more detailed, evidence-based understanding of experts' decision paths and mental models. This will allow us to provide more targeted guidance to less experienced clinicians on how to weigh information.

Third, our study focused on a first implementation; future work is needed to deploy SIT-CARE in a clinical setting and make it available to practicing psychotherapists. However, to do so, extensive onboarding, including detailed information on how to consider specific nonverbal behavior in their reasoning, and a privacy concept, are required. 

Fourth, our underlying AI model of SIT-CARE was trained using data from adult participants in one country. 
This may limit the generalizability of SIT-CARE, because nonverbal behavior is strongly influenced by culture~\cite{gendron2018universality}. Future work should also consider other cultural contexts. 
Furthermore, clinicians typically consider multiple potential diagnoses at once in practice, while SIT-CARE focuses solely on ASC. Thus, considering other diagnoses for which research indicates that nonverbal behavior may deviate from the norm would complement the diagnostic workflow (\eg eating disorders~\cite{caglar-nazaliSystematicReviewMetaanalysis2014, monteleoneNonverbalSocialCommunication2022}, Attention Deficit Hyperactivity Disorder (ADHD) \cite{uekermann2010social}).
However, updated models with different training data, control groups other than Non-ASC, and accuracy levels may be less compatible with clinicians' prior experiences with SIT-CARE and their established current mental model of how the AI may behave (see~\cite{bansal2019updates}). This could negatively impact the diagnostic process. Thus, future versions of SIT-CARE should be developed in collaboration with clinicians with the appropriate backgrounds and expertise.

\section{Conclusion}

Our results demonstrate the potential of SIT-CARE to address key challenges in ASC assessment by integrating clinicians' needs and workflows into the design of an AI-based CDSS. SIT-CARE provides data-driven insights into nonverbal behavior and AI-based recommendations, supporting clinicians across multiple stages of decision-making, from data gathering to final recommendations.
Our study emphasizes the importance of human-centered design in developing AI tools for clinical practice. It reveals that such systems can assist with complex diagnostic tasks and support clinicians' reasoning and decision-making processes. 
Our findings highlight the dual potential of an AI-based CDSS as a screening support tool and a valuable resource for clinician education.
Further, we found that clinicians using SIT-CARE led to four different decision paths, which are reflected in clinicians' decision changes and their mental model such as about the AI's capabilities. 
Future studies should explore ways to integrate SIT-CARE into broader diagnostic workflows and assess its impact in real-world clinical settings.

\section{Statement of use of LLMs}

For this paper, LLMs were not used beyond editing our own text.

\bibliographystyle{ACM-Reference-Format}
\bibliography{bibliography}

\appendix

\section{Formative study interview protocol}
\label{sec:app_interviewscript_1}

Below we provide an outline of the interview phases.

\begin{itemize}
    \item \textbf{Phase 1}: Questions about their professional experience, current work focus, their diagnosis process, and current challenges
    \item \textbf{Phase 2}: Introducing the SIT, Questions about prior knowledge
    \item \textbf{Phase 3}: Open questions about nonverbal behaviors of interest in the SIT context
    \item \textbf{Phase 4}: Open questions about whether nonverbal behaviors of a list are of interest
    \item \textbf{Phase 5}: Exploration and interpretation of schematic visualization of the SIT video data
    \item \textbf{Phase 6}: Question about the possible integration of the SIT into practice
    \item \textbf{Phase 7}: Questions about demographics and concluding information
\end{itemize}

The list of nonverbal behaviors for Phase 4 is provided below.
For gaze behavior the following were listed: Deviation of gaze from the screen center during interaction (time course), Duration of gaze fixation on the screen, Variability of gaze directions, Proportion of gaze directions on the screen, Proportion of gaze directions away from the screen.
For facial expressions the following were listed: Frequency of negative and positive facial expressions, Variability of positive and negative facial expressions, Average intensity of furrowed eyebrows, Intensity of positive facial expressions during interaction (time course).
For head movements the following were listed: Variability of head positions, Frequency of nodding, Duration of individual movements.
For voice the following were listed: Height of vocal pitch, Variability of vocal pitch, Speaking volume, Clarity of voice, Speaking speed.

For phase 5 we provided schematic visualizations via an interactive concept board:

\begin{figure}[!htbp]
    \centering
    \includegraphics[width=1\linewidth]{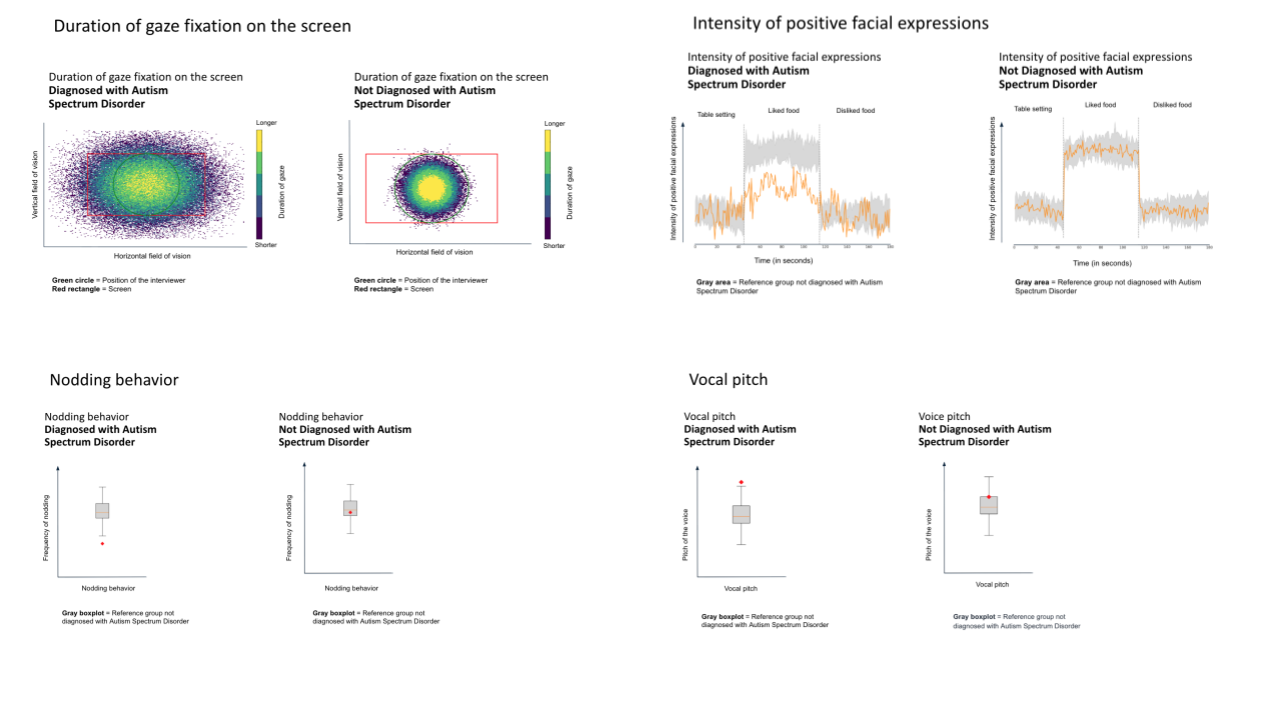}
    \caption{Schematic visualizations used to illustrate how data could be visualized.}
    \Description[<Description of a collage of data visualizations.>]
    {<The image shows visualizations of four data types. For each a schematic example of how the plot could look like for an individual with or without ASC is provided. On the upper left the duration of gaze fixation on the screen is shown as a heatmap. X-axis shows horizontal field of view. Y-Axis shows vertical field of view. A red rectangle shows the screen position and a green circle the interviewer. On the right of each heatmap a legend is shown, indicating the duration of fixation. The Non-ASC example shows a narrowed focus on the interviewed, the ASC example a more diverse fixation. On the upper right, Intensity of positive facial expressions is visualized in a line plot. Y-Axis: Intensity of positive facial expressions, X-Axis: Time (in seconds). The 180 seconds on the X-Axis is split into three parts: Setting the table (neutral), Liked food (positive) and Not liked food (negative). A light grey area shows the reference group not diagnosed with ASC. For the ASC example, the line is shown as lower as the reference group in the Liked food part. For the Non-ASC example, the line matches closer to the reference group. In the down left, Nodding behavior is visualized in a box plot. X-Axis: Nodding behavior. Y-Axis: Frequency of nodding. The gray boxplot represents the reference group not diagnosed with ASC. For the ASC example, the a dot is shown below the box plot quartiles to indicate the hypothetical value of a individual with ASC. For the Non-ASC example, the a dot is shown near the median. In the down right, Vocal pitch is also presented using box plots. X-Axis: Pitch of the voice, Y-Axis: Vocal pitch. The gray boxplot represents the reference group not diagnosed with ASC. The point for vocal pitch for the example with ASC is shown as above the box plot. For the Non-ASC example, the dot is shown near the median.
>}
    \label{fig:placeholder}
\end{figure}

\section{Evaluation study interview protocol}
\label{sec:app_interviewscript_2}

Below we provide an outline of the interview phases. Phase 3 was done twice, first with the Non-ASC case, and then with the ASC case.

\begin{itemize}
    \item \textbf{Phase 1}: Questions about their professional experience with a focus on ASC
    \item \textbf{Phase 2}: Introduction to SIT-CARE
    \item \textbf{Phase 3.1}: Video-based assessment of case 
    \begin{itemize}
        \item Initial assessment of nonverbal behavior with confidence statement
        \item Describe own reasoning
        \item Predict model-based assessment and explain
    \end{itemize}
    \item \textbf{Phase 3.2}: Data-based assessment of case 
    \begin{itemize}
        \item Reevaluate assessment 
        \item Describe own reasoning
        \item Agreement with data-based assessments
        \item Predict model-based assessment and explain
    \end{itemize}
    \item \textbf{Phase 3.3}: Model-based assessment of case 
    \begin{itemize}
        \item Reevaluate assessment 
        \item Describe own reasoning
    \end{itemize}
    \item \textbf{Phase 4}: Questions about comprehension, predictability, helpfulness and trustworthiness of outputs
    \item \textbf{Phase 5}: Questions about what they liked, disliked and would like to improve
    \item \textbf{Phase 6}: Questions about integration into workflow
    \item \textbf{Phase 7}: Attitudes towards AI
\end{itemize}

\section{Data-based visualization}
\label{sec:app_visualizations}

\begin{figure}[H]
    \centering
    \includegraphics[width=0.75\linewidth]{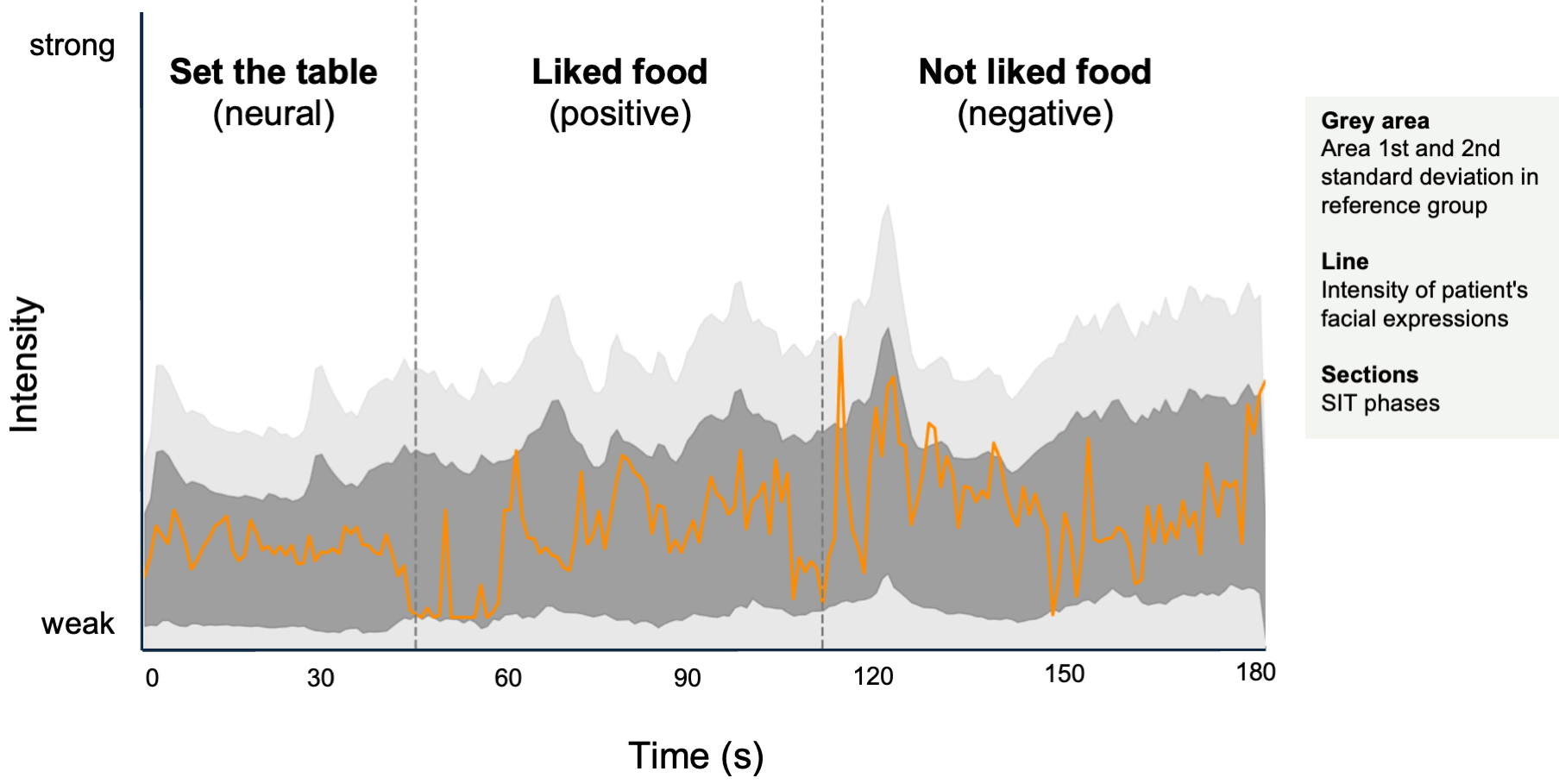}
    \caption{Visualization of the chosen facial expression, intensity and variability of positive facial expressions per phase, of a hypothetical Non-ASC patient.}
    \label{fig:facial}
    \Description[<Description of a line plot visualizing facial expression data.>]
    {<The graph plots the intensity of a hypothetical patient's facial expression over time the duration of the SIT across three SIT phases: Setting the table (neutral), Liked food (positive) and Not liked food (negative). The x-axis represents time in seconds ranging from zero to 180. The y-axis represent intensity of the facial expressions ranging from weak on the bottom to strong on the top. To prevent overinterpretation, no numerical values were provided on the y-axis. A colored line represents the patient's expression intensity over time. A shaded area indicates the first and second standard deviations within the reference group. The patient displayed less intense expressions in neutral situations than in positive or negative ones. The strongest expression intensity occurred during the negative phase, where the patient's line rises above the 1st area of the reference group. On the right is a legend describing what the line, shaded area and phases are.>}
\end{figure}

\begin{figure}[H]
    \centering
    \includegraphics[width=0.75\linewidth]{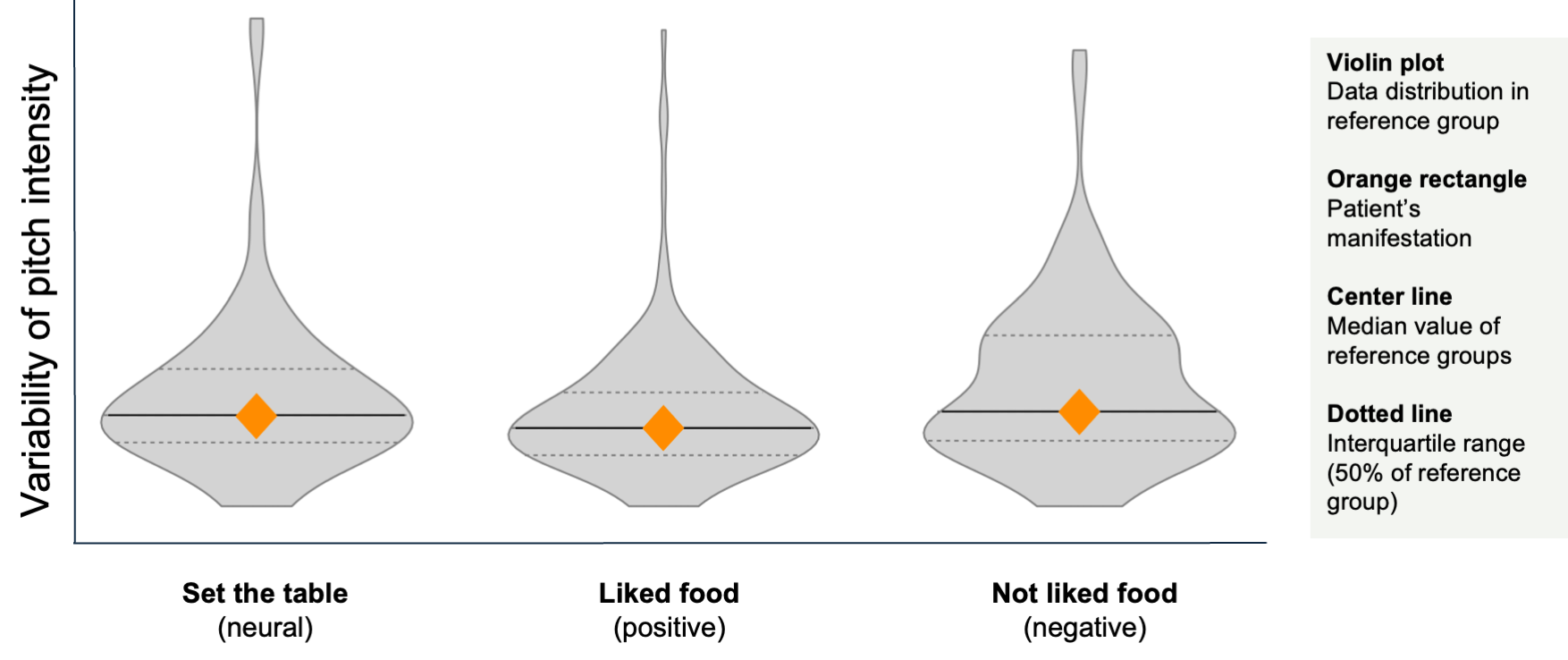}
    \caption{Visualization of the chose voice data variability of pitch intensity per phase, of a hypothetical Non-ASC patient.}
    \label{fig:voice}
    \Description[<Description of three violin plots visualizing voice data.>]
    {<Three violin plots visualize the variability of pitch intensity by SIT phase. The x-axis represents the three phases Setting the table (neutral), Liked food (positive) and Not liked food (negative). For each phase one violin plot is shown. The y-axis represent variability of pitch intensity. To prevent overinterpretation, no numerical values were provided on the y-axis. The hypothetical patient's values align with the median of the reference group. Neutral and positive phases are similar, while the negative phase shows greater variance and higher pitch intensity.>}
\end{figure}

\section{Detailed calculations for data-based assessment}
\label{app:calculations}

This appendix provides the detailed computational procedures for \autoref{subsec:implementation_backend}. 

\subsection{Gaze behavior}
\paragraph{Fixation density levels.}  
To create the bi-dimensional histogram plots of gaze fixations, raw gaze vectors were projected from the webcam coordinate system to a screen-centered coordinate space. Fixation density was color-coded into five discrete levels, defined relative to a reference group (RG, Non-ASC participants) as:

\[
\text{Intervals}_{\text{RG}} = \left\{
\begin{array}{ll}
(-\infty,\ \mu_{\text{RG}} - 1.5\sigma_{\text{RG}}) & \text{very low fixation density} \\
\left[\mu_{\text{RG}} - 1.5\sigma_{\text{RG}},\ \mu_{\text{RG}} - 0.5\sigma_{\text{RG}}\right) &  \\
\left[\mu_{\text{RG}} - 0.5\sigma_{\text{RG}},\ \mu_{\text{RG}} + 0.5\sigma_{\text{RG}}\right) &  \\
\left[\mu_{\text{RG}} + 0.5\sigma_{\text{RG}},\ \mu_{\text{RG}} + 1.5\sigma_{\text{RG}}\right) & \\
(\mu_{\text{RG}} + 1.5\sigma_{\text{RG}},\ \infty) & \text{very high fixation density}
\end{array}
\right.
\]

where \(\mu_{\text{RG}}\) and \(\sigma_{\text{RG}}\) denote the mean and standard deviation of fixation counts in the reference group.  

\paragraph{Representative examples.}  
To illustrate prototypical gaze distributions, we pooled all frame-level gaze points for ASC and Non-ASC groups separately. From each pool, we randomly sampled 5,500 frames (approx.\ equal to the number of frames per participant) using a fixed random seed (\texttt{random\_state=42}) to ensure reproducibility. These sampled frames were then visualized as bi-dimensional histogram plots.

\paragraph{Numerical statistics.}  
In addition to visualizations, the following statistics were computed for each assessed participant:
\begin{enumerate}
    \item Variance of horizontal gaze angles.
    \item Percentage of time spent looking at the screen.
    \item Percentage of time directed toward the actress’s face region.
\end{enumerate}
For comparison, mean and standard deviation of these metrics were computed for both ASC and Non-ASC reference groups.

\subsection{Facial expressions}
\paragraph{Representative examples.}  
Representative individuals were identified using a distance-to-median score:

\[
\text{score}_i
= \frac{\text{total\_abs\_diff}_i}
     {\operatorname{median}_j(\text{total\_abs\_diff}_j)}
+
\sum_{p \in \{\text{neutral, joy, disgust}\}}
\frac{\left|\operatorname{Var}_i^{(p)} - \operatorname{median}_j\!\left(\operatorname{Var}_j^{(p)}\right)\right|}
     {\operatorname{median}_j\!\left(\left|\operatorname{Var}_j^{(p)} - \operatorname{median}_k\!\left(\operatorname{Var}_k^{(p)}\right)\right|\right)} .
\]

Here, \(\mathrm{total\_abs\_diff}_i\) is the sum of absolute deviations between subject \(i\)’s trace and the group median trace, while \(\operatorname{Var}_i^{(p)}\) denotes variance within phase \(p\). The two subjects with the lowest scores were visualized as prototypical examples. For anonymization, 15\% Gaussian noise was added to the traces.  

\paragraph{Numerical statistics.}  
For each participant, we computed:
\begin{enumerate}
    \item Mean and variance of smiling intensity (AU06 + AU12).
    \item Mean and variance of AU04 (brow lowering, associated with negative affect).
    \item Overall mean intensity across all AUs.
\end{enumerate}
Reference distributions (mean \(\pm\) SD) were calculated separately for ASC and Non-ASC groups, and additionally for gender-specific subgroups.

\subsection{Voice parameters}
For voice features, we extracted from the openSMILE eGeMAPS set:
\begin{enumerate}
    \item Variance of pitch (\texttt{F0semitoneFrom27.5Hz\_sma3nz\_pctlrange0-2}) (representing the range of pitch (in semitones, relative to 27.5 Hz) between the 2nd and 98th percentiles of voiced segments, thereby minimizing the effect of outliers).
    \item Mean and variance of voiced segment length (\texttt{MeanVoicedSegmentLengthSec}, \texttt{StddevVoicedSegmentLengthSec$^2$}).
    \item Loudness variability (\texttt{loudness\_sma3\_stddevNorm$^2$}).
\end{enumerate}

Prototypical examples corresponded to the median values of ASC and Non-ASC groups.
Group-level means and standard deviations were computed for ASC and Non-ASC reference groups. Visualizations of pitch variability were provided via violin plots, stratified by interaction phase, while numerical results were displayed alongside the assessed participant’s values.


\end{document}